\documentclass[12pt]{iopart}
\usepackage[scr=boondoxo,scrscaled=1.05]{mathalfa}
\usepackage{tikz}
\usetikzlibrary{calc,decorations.markings}
\usepackage{epsfig}
\usepackage{graphicx}
\usepackage{rotating}
\expandafter\let\csname equation*\endcsname\relax
\expandafter\let\csname endequation*\endcsname\relax
\usepackage{amsmath,amssymb}
\usepackage[a4paper]{geometry}
\usepackage{subcaption}
\usepackage{float} 
\usepackage[font=small,labelfont=bf,justification=justified,format=plain]{caption}
\usepackage[active]{srcltx}
\usepackage{url}
\usepackage{centernot}
\usepackage{dsfont,bm}
\usepackage{pgfplots}
\usepackage{stackengine}
\usepackage{enumitem}
\usepackage[scr=boondoxo,scrscaled=1.05]{mathalfa}
\usepackage{hyperref}
\colorlet{darkred}{red!85!black}
\colorlet{darkgreen}{green!50!black}
\colorlet{darkblue}{blue!60!black}
\hypersetup{
	colorlinks   = true, 
	urlcolor     = darkgreen, 
	linkcolor    = darkblue, 
	citecolor   = darkred 
}
\usepackage{tikz}
\usepackage{tkz-euclide}
\usetikzlibrary{calc,decorations.markings}
\usepackage{times,graphicx,xcolor}
\usepackage{delarray,fancybox}
\usepackage{mathdots}
\usepackage{eurosym}
\usepackage{relsize}
\usepackage{soul}

\newtheorem{proposition}{\bf Proposition}[section]
\newtheorem{definition}{\bf Definition}[section]

\colorlet{darkred}{red!85!black}
\colorlet{darkgreen}{green!50!black}
\colorlet{darkblue}{blue!60!black}
\definecolor{newred}{RGB}{125,0,45}
\definecolor{newblue}{RGB}{0,80,158}
\definecolor{newgray}{gray}{0.4}
\definecolor{ArmyGreen}{rgb}{0.29, 0.33, 0.13}
\definecolor{BostonRed}{rgb}{0.9, 0.0, 0.0}
\definecolor{CadmiumGreen}{rgb}{0.0, 0.42, 0.24}
\definecolor{darkyellow}{RGB}{205,205,0}
\definecolor{Gray}{gray}{0.9}
\definecolor{LightCyan}{rgb}{0.88,1,1}
\definecolor{PastelGreen}{rgb}{0.47,0.87,0.47}
\definecolor{PastelRed}{rgb}{0.87,0.47,0.47}
\definecolor{PastelYellow}{rgb}{0.99, 0.99, 0.59}
\definecolor{platinum}{rgb}{0.9, 0.89, 0.89}
\definecolor{pistachio}{rgb}{0.58, 0.77, 0.45}
\definecolor{princetonorange}{rgb}{1.0, 0.56, 0.0}
\definecolor{FigureBlue}{rgb}{0.447, 0.624, 0.812}
\definecolor{NewBlue}{rgb}{0.712, 0.806, 0.861}
\definecolor{LightBlue}{rgb}{0.812, 0.906, 0.961}
\definecolor{airforceblue}{rgb}{0.36, 0.54, 0.66}
\definecolor{amber}{rgb}{1.0, 0.75, 0.0}

\def\<{\langle}
\def\>{\rangle}

\newcommand{\cab}[2]{\mathbf{e}_{#1}^{\scriptscriptstyle{(#2)}}}
\newcommand{\cad}[2]{\mathbf{e}_{#1}^{\scriptscriptstyle{(#2)}}{}^{\dagger}}

\newcommand{\smi}[1]{\scriptscriptstyle{(#1)}}

\begin{document}
	
	\title{Universal bound on the Lyapunov spectrum of quantum master equations}
	
	\author{Paolo Muratore-Ginanneschi $ ^1 $, Gen Kimura $ ^2$, Frederik vom Ende $^3$ and Dariusz Chru\'sci\'nski $ ^4$}
	
	\address{$^1$ University of Helsinki, Department of Mathematics and Statistics
		P.O. Box 68 FIN-00014, Helsinki, Finland. \\
	} 
	\vspace{-0.2cm}
	\ead{paolo.muratore-ginanneschi@helsinki.fi}
	\vspace{0.4cm}
	\address{
		$^2$ Graduate School of Information Sciences, Tohoku University	Sendai, 980-8579, Japan. \\
	} 
	\vspace{-0.2cm}
	\ead{gen@shibaura-it.ac.jp}
	\vspace{0.4cm}
	\address{
		$^3$ Dahlem Center for Complex Quantum	Systems, Freie Universit\"at Berlin, Arnimallee 14, 14195 Berlin, Germany. \\
	}	
	\vspace{-0.2cm}
	\ead{frederik.vomende@gmail.com}
	\vspace{0.4cm}
	\address{
		$ ^4 $ Institute of Physics, Faculty of Physics, Astronomy and Informatics, Nicolaus Copernicus University, Grudziadzka 5/7, 87-100 Toru\'n, Poland.
	}
	\vspace{-0.2cm}	
	\ead{darch@fizyka.umk.pl}
\begin{abstract}
The spectral properties of positive maps are pivotal for understanding the dynamics of quantum systems interacting with their environment. Furthermore, central problems in quantum information such as the characterization of entanglement can be reformulated in terms of spectral properties of positive maps. The present work aims to contribute to a better understanding of the spectrum of positive maps. Specifically, our main result is a new proof of a universal bound on the $d^{2}-1$ generically non vanishing decay rates $\Gamma_{i}$ of time-autonomous quantum master equations on a $d$-dimensional Hilbert space:
$$\phantom{aaaa}\Gamma_{\mathrm{max}}\,\leq\,\varkappa_{d}\,\sum_{i=1}^{d^{2}-1}\Gamma_{i}$$
The prefactor $\varkappa_{d}$ 
depends only on the dimension $d$ and 
varies depending on the sub-class of positive maps to which the semigroup solution of the master equation belongs. We provide a brief but self-consistent survey of these concepts. We obtain our main result by resorting to the theory of Lyapunov exponents, a central concept in the study of dynamical systems, control theory, and out-of-equilibrium statistical mechanics. We thus show that progress in understanding positive maps in quantum mechanics may require ideas at the crossroads between different disciplines. For this reason, we adopt a notation and presentation style aimed at reaching readers with diverse backgrounds.
\end{abstract}

\section{Introduction}

Physics is an experimental science. A recurring question is how to extract dynamical laws from experimental data.
In the case of open quantum systems, this question is often formulated by asking whether an observed evolution (dynamical map) can be mathematically understood as a family of linear maps that enjoys the semigroup property and whose members are all positivity preserving. In that case, there also exists a precise and computable answer for time-autonomous dynamics \cite{WoEiCuCi2008}. Specifically, one needs to check whether the logarithm of a tomographically estimated dynamical map at a given time, i.e., a single snapshot of the dynamical map, satisfies the necessary and sufficient conditions characterizing generators of completely positive semigroups and, therefore, solutions of the Lindblad \cite{LinG1976} Gorini-Kossakowski-Sudarshan \cite{GoKoSu1976} master equation.
These conditions are preservation of the trace and self-adjointness, as well as conditional complete positivity \cite{EvaD1977} (see also \cite[\S~7]{WolM2012}) of the generator. Whilst the answer is mathematically complete and constructive, actual implementations encounter several stumbling blocks. To start with, the complex logarithm is a multi-valued function so that, in general, checks may require to extend the search to multiple branches. In addition, the search must be robust against state preparation and measurement errors that occur in realistic situations. Finally, the computational cost grows exponentially with the number of dimensions. Connecting to this cost, it is also worth mentioning that it is in fact possible to prove \cite{CuEiWo2012} that the $1$-IN-$3$-SAT, an instance of a non-deterministic polynomial time complete (i.e. NP-complete) problem in computational complexity theory \cite[\S~5]{MoMe2011}, is amenable by a Turing reduction to proving that the exponential of a certain non-representational (i.e., abstract and non related to any physical model \cite{IpCa2021}) generator yields an element of a completely positive semigroup. This means that \emph{in principle} assessing whether the logarithm of a dynamical map specifies the generator of a completely positive semigroup is at least as hard as solving any NP problem on a classical computer.
In practice and for systems with few degrees of freedom, the result of \cite{WoEiCuCi2008} is the basis for computationally viable algorithms 
which are relevant for to-date quantum computing hardware \cite{LiSeKoOnCu2025}.

The difficulties discussed above motivate the interest for partial but practically more direct answers that provide necessary conditions (``witnesses'') to identify a completely positive semigroup. It was conjectured long ago \cite{KimG2002} that complete positivity of a homogeneous semigroup of dynamical maps necessarily entails an upper bound on the relaxation rates characterizing the contractive forward evolution. Relaxation rates are commonly measured in laboratory experiments. The conjecture was later extensively corroborated by proofs under additional hypotheses and examples exhibiting tightness of the putative bound \cite{ChKiKoSh2021}. Mathematically, relaxation rates are the real parts of the logarithms of the eigenvalues of the flow solution of a linear differential equation. If the differential equation is time-autonomous, the forward in time flow it generates constitutes a homogeneous semigroup and the relaxation rates coincide with the Lyapunov exponents \cite{AdrL1995}. These considerations led to a first proof of the conjecture by combining ideas of dynamical systems and open quantum system theory \cite{MGKiCh2025}. The proof, however, left open the question of tight bounds for the relaxation rates of positive but non completely positive maps. 
This is an important question because positive (i.e., positivity-preserving) linear maps---which are fundamental to quantum mechanics---have a far more articulated structure than those appearing in the theory of classical stochastic processes \cite{ChKo2009}. 
In \cite{ChvEnKiMG2025,vEnChKiMG2025} we were able to provide a complete answer to this question using different and more traditional methods of algebraic quantum theory. Specifically, the results of \cite{ChvEnKiMG2025,vEnChKiMG2025} show that the same bound holds for maps that are at least $2$-positive and that this requirement cannot be weakened further. We emphasize that $2$-positive maps are completely positive maps for single-qubit systems. The relaxation rates of simply positive (or $1$-positive) maps only obey the trivial bound dictated by contractivity, as do classical positive maps. Finally, our results in \cite{ChvEnKiMG2025,vEnChKiMG2025} single out the class of Schwarz maps, a subset of positive maps that contains the class of $2$-positive maps: we show that the relaxation rates of Schwarz maps obey a weaker yet non trivial bound than that of $2$-positive maps.

The scope of the present contribution is to detail an alternative proof of the results of \cite{ChvEnKiMG2025,vEnChKiMG2025}
by revisiting the connection with Lyapunov exponents in dynamical systems theory. We believe this is interesting for several reasons.
The advent of nano-technologies has given a lot of impulse to the study of finite-time (stochastic) thermodynamics, thus attracting
to the theory of open quantum systems also scholars with multifarious backgrounds and, in particular, statistical physics. The derivation of the results of \cite{ChvEnKiMG2025,vEnChKiMG2025} by directly relating the properties of positive maps in quantum theory with Lyapunov exponents---dynamical systems indicators of cross-disciplinary interest (see, e.g., \cite{CeCeVu2009,PiPo2016,MaShSt2016})---may be useful to a wider community than the one traditionally interested in open quantum systems. 
In addition, we also show how to use the affine invariance of Lyapunov exponents and the Lozinskii-Dahlquist 
estimates to derive our main results \cite{LozS1958,DahG1959} (see also \cite[\S~IV.6]{AdrL1995}). These are general results to which the current literature on Lyapunov exponents often pays little attention to (see, however, the recent \cite{AmSo2014,CiVeJaBu2022}), probably as a consequence of the general appeal for computational applications of formulations based on Oseledets' multiplicative ergodic theorem \cite{OseV1968,BeGaGiSt1980,BeGaGiSt1980a}. 
Finally, our results provide tight bounds on the maximal Lyapunov exponent of a dynamical system, hence accomplishing a task generically known to be NP-hard \cite{TsBl1997}. 

With these considerations in mind, we have chosen a presentation style and notation (e.g., linear algebra rather than Dirac's bra-ket notation from quantum physics) that is likely to be more familiar to readers whose principal research interests lie, for example, in dynamical systems, control theory, and statistical physics.

The structure of the paper is as follows. In section~\ref{sec:LPM} we briefly recall basic facts about positive linear maps in classical and quantum physics. In section~\ref{sec:ECME} we recall that any generator of trace- and self-adjointness preserving linear maps embeds a classical master equation \cite{TeMa1992} if and only if the Kossakowski conditions hold \cite{KosA1972,GoKoSu1976} (see also \cite[\S~4.3]{RiHu2012}). In section~\ref{sec:conditions} we prove that the foliation of the set of positive maps in quantum mechanics in a hierarchy of nested subsets translates into conditions on the contraction rate of the volume form of state operators. These results are instrumental to prove the main result of this work. In section~\ref{sec:Lyapunov} we briefly review basic concepts of the theory of Lyapunov exponents that we use in the proof of our main result. We give particular emphasis to Lyapunov regularity, a property that quantum master equations relevant for applications, most often time-autonomous or time-(almost)-periodic, enjoy. Section~\ref{sec:main} contains our main result. We include some examples illustrating the bounds in section~\ref{sec:examples}. The last section is devoted to conclusions and outlook. 

\section{Positive maps solving differential equations}
\label{sec:LPM}

We first recall some facts about classical master equations and how they generate semigroups of positive maps. We then turn to the quantum case to very briefly summarize analogies and differences. Readers familiar with the topics may want to skip this section.

\subsection{Classical master equations}

Probability vectors valued on a finite-dimensional configuration space are the main objects in the theory of classical stochastic processes. These are real vectors with positive entries that sum to one. Geometrically, they form a simplex, a compact convex subset of a real Euclidean vector space \cite[\S~2]{BeZy2006}.

If the stochastic process is a differentiable Markov chain on a $d$-dimensional Euclidean configuration space, then the family of probability vectors $\mathsf{p}(t) $, that describes the state of the process as $t$ increases, satisfies a first-order system of differential equations of the form
\begin{align}
	\dot{\mathsf{p}}(t)=\mathsf{K}(t)\,\mathsf{p}(t)\,.
	\label{LPM:cme}
\end{align}
We refer to the family of $d\,\times\,d$-matrices $\mathsf{K}(t)$ as the (time-dependent) generator of the dynamics.
We require $\mathsf{K}(t)$ to depend on the time variable $t$ in a sufficiently regular manner to satisfy the conditions of global existence of solutions \cite[\S~1]{AdrL1995}. In addition, we require $ \mathsf{K}(t)$ to be column-diagonally dominated with positive off-diagonal entries, that is,
\begin{align}
	\mathsf{K}_{i,j}(t)
	\begin{cases}
		\,\geq\,0& \forall\,i\neq j
		\\[0.3cm]
		=-\sum_{k\neq i}^{d} \mathsf{K}_{k,i}(t)	&\forall\, i=j
	\end{cases}
	\label{LPM:cgen}
\end{align}
Under these conditions, (\ref{LPM:cme}) specifies a classical master equation governing the forward in time evolution of probability vectors of the Markov chain. In particular, for each $t\,\geq\,s$ the matrix elements of the time-ordered exponential
\begin{align}
	\mathsf{T}(t,s)=\mathcal{T}\exp\left(\int_{s}^{t}\mathrm{d}u\,\mathsf{K}(u)\right)
	\label{LPM:tp}
\end{align}
solving (\ref{LPM:cme}) with initial condition specified by the identity matrix in $\mathcal{M}_{d}(\mathbb{C})$
\begin{align}
		\mathsf{T}(s,s)=\mathsf{1}_{d}
\label{LPM:init}
\end{align}
specify the transition probabilities to go from any state at the earlier time $s$ to another state at the later time $t$. The theory of differential equations guarantees that the maps (\ref{LPM:tp}) collectively enjoy the semigroup property
\begin{align}
\mathsf{T}(t_{3},t_{1})=\mathsf{T}(t_{3},t_{2})\mathsf{T}(t_{2},t_{1}) \qquad \forall\,t_{3}\,\geq\,t_{2}\,\geq\,t_{1}
	\label{LPM:sg}
\end{align}
which guarantees that the Chapman-Kolmogorov equation, a necessary consequence of the Markov property, is satisfied \cite[\S~2]{PavG2014}. This means that the transition probability matrix (\ref{LPM:tp}) maps the simplex of probability vectors in $\mathbb{R}^{d}$ into itself. We thus say that (\ref{LPM:tp}) defines a family of classical positive linear maps.
A further direct consequence of the interpretation of (\ref{LPM:tp}) as transition probabilities is that the dynamics
generated by (\ref{LPM:cme}) is contractive with respect to the $1$-norm
\begin{align}
	\|\mathsf{T}(t,s)(\mathsf{p}(s)-\tilde{\mathsf{p}}(s))\|_{1}
	\,\leq\,\left\|\mathsf{p}(s)-\tilde{\mathsf{p}}(s)\right\|_{1}\,\equiv\,\sum_{i=1}^{d}|\mathsf{p}_{i}(s)-\tilde{\mathsf{p}}_{i}(s)|\,.
	\label{LPM:1n}
\end{align}
Combining contractivity with the property of mapping a convex compact set into itself paves the way for the application of Perron-Frobenius theorem (see e.g. \cite[\S~1]{SenE2006}). The upshot is that the spectrum of any of the elements of the semigroup (\ref{LPM:tp}) is confined to the unit disk of the complex plane, and it is symmetric with respect to the real axis. Furthermore, there is at least one eigenvalue equal to one. The right eigenoperator associated with this eigenvalue is a state operator after appropriate normalization. 

\subsection{Quantum master equations}

State operators are the main objects in the theory of open quantum systems. In the case of a Hilbert space with $d$ dimensions, they are positive semi-definite (henceforth just called ``positive'') $d\,\times\,d$ matrices with unit trace. Geometrically, they also form a compact and convex subspace of the matrix space $\mathcal{M}_{d}(\mathbb{C})$ of $d\,\times\,d$ matrices with complex entries \cite[\S~8.4]{BeZy2006}.

 Based only on the requirements of trace and self-adjointness preservation, it is in general possible to derive the matrix valued equation
\begin{align}
	\dot{\bm{\rho}}(t)=\mathfrak{L}(t)[\bm{\rho}(t)]\,.
	\label{LPM:qme}
\end{align}
The action of the generator $\mathfrak{L}(t)$ on any $\operatorname{O} \in \mathcal{M}_{d}(\mathbb{C})$ is always amenable to the following canonical form (see, e.g., \cite{HaCrLiAn2014})
\begin{align}
	\begin{split}
	&\mathfrak{L}(t)[\operatorname{O}]=-\imath\,\left[\operatorname{H}(t),\operatorname{O}\right]+
	\frac 12 \sum_{\ell=1}^{d^{2}-1}\mathscr{c}_{\ell}(t) \Big( \left[\operatorname{L}_{\ell}(t)\,,\operatorname{O}\operatorname{L}_{\ell}^{\dagger}(t)\right]+\left[\operatorname{L}_{\ell}(t)\operatorname{O}\,,\operatorname{L}_{\ell}^{\dagger}(t)\right]\Big) 
	\\
	&	\begin{cases}
		\operatorname{Tr}\left(\operatorname{L}_{\ell}^{\dagger}(t)\operatorname{L}_{k}(t)\right)=\delta_{\ell,k}	
		\\
		\operatorname{Tr}\left(\operatorname{L}_{\ell}(t)\right)=0		
		\\
		\operatorname{H}(t)=\operatorname{H}^{\dagger}(t)
	\end{cases}
	\hspace{0.5cm} \forall\,t\,\geq\,0 \hspace{0.5cm}	\&\hspace{0.5cm} \forall\,\ell,k=1,\dots,d^{2}-1			
	\end{split}
	\label{LPM:gen}
\end{align} 
where $\operatorname{H}(t) $ is a Hamilton operator, the $ \mathscr{c}_{\ell}(t)$'s are scalar couplings, and the $\operatorname{L}_{\ell}(t)$'s are commonly referred to as decoherence or jump operators. As in the classical case, here we always assume that any explicit time dependence is sufficiently regular, e.g. continuous and uniformly bounded, to satisfy the conditions for global existence of solutions of (\ref{LPM:qme}) for any positive time $t\,\geq\,0$.  

We refer to (\ref{LPM:qme}), always implying the generator form (\ref{LPM:gen}), as {\textquotedblleft}quantum master equation{\textquotedblright}. The main problem with (\ref{LPM:qme}) is that it does not ensure the preservation of positivity without further assumptions on the scalar couplings $ \mathscr{c}_{\ell}(t)$.

A cornerstone result of the theory of open quantum systems is the proof, see e.g. \cite[Theorem~4.2.1]{RiHu2012}, that if and only if
\begin{align}
\mathscr{c}_{\ell}(t)\,\geq\,0 \qquad\forall\,\ell=1,\dots,d^{2}-1
	\label{LPM:LGKS}
\end{align}
then solutions of (\ref{LPM:qme}) can always be couched in the readily positivity-preserving operator sum form (often referred to as the Kraus form \cite{KraK1983})
\begin{align}
\Phi(t,s)[\bm{\rho}(s)]=\sum_{\ell=0}^{d^{2}-1}\operatorname{V}_{\ell}(t,s)\bm{\rho}(s)\operatorname{V}_{\ell}^{\dagger}(t,s) \qquad \forall\,t\,\geq\,s
	\label{LPM:povm}
\end{align}
The collection of matrices $\operatorname{V}_{\ell}(t,s)\in \mathcal{M}_{d}(\mathbb{C})$ is such that (\ref{LPM:povm}) reduces to the identity map for $t=s$ and
\begin{align}
\sum_{\ell=0}^{d^{2}-1}\operatorname{V}_{\ell}^{\dagger}(t,s)\operatorname{V}_{\ell}(t,s)=\operatorname{1}_{d} \qquad \forall\,t\,\geq\,s
\label{LPM:complete}
\end{align}
to ensure trace preservation. 

	The proof of (\ref{LPM:LGKS})-(\ref{LPM:povm}) was first obtained in groundbreaking papers by Lindblad \cite{LinG1976} and Gorini-Kossakowski-Sudarshan \cite{GoKoSu1976} under the hypothesis of a time-autonomous generator. This means the case where in (\ref{LPM:gen}) the Hamilton operator $\operatorname{H}$, the scalar couplings $ \mathscr{c}_{\ell}$'s, and decoherence operators $\operatorname{L}_{\ell}$'s are all time independent. In the time-autonomous case (\ref{LPM:LGKS}) straightforwardly reduces to the requirement that the $ \mathscr{c}_{\ell}$'s are positive real numbers:
    \begin{equation}
    \dot{\bm{\rho}}(t) =   \mathfrak{L}[\bm{\rho}(t)] :=-\imath\,\left[\operatorname{H},\bm{\rho}(t)\right]+
	\frac 12 \sum_{\ell=1}^{d^{2}-1}\mathscr{c}_{\ell}\Big( \left[\operatorname{L}_{\ell}\,,\bm{\rho}(t)\operatorname{L}_{\ell}^{\dagger}\right] + \left[\operatorname{L}_{\ell} \bm{\rho}(t)\,,\operatorname{L}_{\ell}^{\dagger}\right]\Big) , \ \ c_\ell \geq 0 .
\end{equation}
    As a consequence, solutions of the master equation only depend upon the time elapsed from the moment an initial state operator $\bm{\rho}$ is assigned:
    \begin{equation}
        \bm{\rho}(t) = \Phi(t,0)[\bm{\rho}],
    \end{equation}
    and
	\begin{align}
		\Phi(t,s) = \Phi(t-s,0) = e^{(t-s) \mathfrak{L}}. 
		\label{LPM:tti}
	\end{align} 
	Correspondingly, we can regard the collection of linear maps $\Phi$ and matrices $\operatorname{V}_{\ell}$'s as forming a one parameter family, also called homogeneous, with respect to the time variable. The time-autonomous case is of particular interest for applications  because it describes quantum master equations obtained in the van-Hove scaling limit \cite{DavE1974}.

It is worth recalling that we can always couch the matrix equation (\ref{LPM:qme}) in the form of a vector equation using the vectorization isomorphism which relates the outer and tensor products \cite[\S~8.2]{BeZy2006}. Accordingly, a solution of (\ref{LPM:qme}) is in a one-to-one correspondence with a family of vectors in $\mathbb{C}^{d^{2}} = \mathbb{C}^{d} \otimes \mathbb{C}^{d}$ 
\begin{align}
\bm{\rho}(t)=	\sum_{i,j=1}^{d}\cab{i}{d}\cad{j}{d}\rho_{i,j}(t)\quad \Longleftrightarrow \quad \mathscr{r}(t)=\sum_{i,j=1}^{d}\cab{i}{d}\otimes\cab{j}{d} \rho_{i,j}(t) 
\label{LPM:reshape}
\end{align}
where $\left\{ \cab{i}{d} \right\}_{i=1}^{d}$ is the canonical (or computational) basis of $\mathbb{C}^{d}$. In other words, (\ref{LPM:reshape}) says that we can construct a vector from a matrix by stacking rows one above the other. 
Upon applying this prescription and using the overline symbol $\,\overline{(\cdot)}\,$ to denote entry-wise complex conjugation, the evolution map (\ref{LPM:povm}) takes the familiar form in linear dynamical systems theory (see, e.g., \cite{CeCeVu2009}) of a matrix acting on a vector:
\begin{align}
	\mathscr{F}(t,s)\mathscr{r}(s)=\left(\sum_{\ell=0}^{d^{2}-1}\operatorname{V}_{\ell}(t,s)\otimes\overline{\operatorname{V}_{\ell}}(t,s) \right)\mathscr{r}(s)\qquad \forall\,t\,\geq\,s
	\label{LPM:vectorialize}
\end{align}
For any fixed values of $t$ and $s$, $ \mathscr{F}(t,s)$ is a $ d^{2}\,\times\,d^{2}$ complex square matrix: 
\begin{align}
	 \mathscr{F}(t,s)\in\mathcal{M}_{d^{2}}(\mathbb{C}) = \mathcal{M}_{d}(\mathbb{C}) \otimes \mathcal{M}_{d}(\mathbb{C})
\end{align} 
 As we vary $t$ and $s$, the $ \mathscr{F}(t,s)$'s specify the elements of the semigroup solving the {\textquotedblleft}\emph{reshaped}{\textquotedblright} master equation \cite[\S~10.2]{BeZy2006}
 \begin{align}
 	\begin{split}
 		& 	\partial_{t}\mathscr{F}(t,s)=\mathscr{L}(t)\mathscr{F}(t,s)
 		\\
 		& 	\mathscr{F}(s,s)=\mathsf{1}_{d^{2}}
 	\end{split}
\qquad\forall \,t\,\geq\,s
 	\label{LPM:qds}
 \end{align}
Here, $ \mathsf{1}_{d^{2}}$ is the identity matrix in $\mathcal{M}_{d^{2}}(\mathbb{C})$  and
 \begin{align}
 	\mathscr{L}(t)&=-\imath\,\Big{(}\operatorname{H}(t)\otimes\mathsf{1}_{d}-\mathsf{1}_{d}\otimes\operatorname{H}^{\top}(t)\Big{)}
 	\nonumber\\
 	&+\sum_{\ell=1}^{d^{2}-1}\mathscr{c}_{\ell}(t)\left(\operatorname{L}_{\ell}(t)\otimes\overline{\operatorname{L}_{\ell}}(t)-\frac{\operatorname{L}_{\ell}^{\dagger}(t)\operatorname{L}_{\ell}(t)\otimes\mathsf{1}_{d}+\mathsf{1}_{d}\otimes\operatorname{L}_{\ell}^{\top}(t)\overline{\operatorname{L}_{\ell}}(t)}{2}\right)
 	\label{LPM:reshaped}
 \end{align}
 We use the superscript $\top$ to denote matrix transposition. The one-to-one correspondence between (\ref{LPM:qds}) and (\ref{LPM:LGKS}) implies that the operator sum (\ref{LPM:povm}) also enjoys the semigroup property:
\begin{align}
\Phi(t_{3},t_{2})\big{[}\Phi(t_{2},t_{1})[\operatorname{O}]\big{]}=\Phi(t_{3},t_{1})[\operatorname{O}]
\qquad \forall\,t_{3}\,\geq\,t_{2}\,\geq\,t_{1} \quad\&\quad \forall\,\operatorname{O}\in\mathcal{M}_{d}(\mathbb{C})
\end{align} 
The semigroup generated by the quantum master equation (\ref{LPM:qme}) subject to the conditions (\ref{LPM:LGKS}) always preserves the state operators. All these facts combined are reminiscent of the properties of semigroups generated by the classical master equation (\ref{LPM:cme}). For this reason quantum master equations satisfying (\ref{LPM:LGKS}) are often called {\textquotedblleft}Markovian{\textquotedblright}. The set of positive maps in quantum theory is, however, much more involved than in the classical case. The positive operator sum form (\ref{LPM:povm}) is only a \emph{sufficient} condition for positivity preservation. In particular,  (\ref{LPM:povm}) is equivalent to adding to positivity preservation the requirement \cite{StiW1955,ChoM1975} that the map on $\mathcal{M}_{d^{2}}(\mathbb{C})$ obtained by the tensor product of (\ref{LPM:povm}) with the identity on $\mathcal{M}_{d}(\mathbb{C})$ is also positive:
\begin{align}
\tilde{\Phi}^{(d)}(t,s)[\tilde{\operatorname{O}}]:=(\mathrm{Id}_{d}\otimes\Phi(t,s))[\tilde{\operatorname{O}}]\,\geq\,0
\qquad \forall\,t\,\geq\,s
	\label{LPM:CP}
\end{align}
for all positive matrices $\tilde{\operatorname{O}}$ in $\mathcal{M}_{d^{2}}(\mathbb{C})$. The condition (\ref{LPM:CP}) is called $d$- or complete positivity. As a consequence, the master equation (\ref{LPM:qme}) subject to the Lindblad-Gorini-Kossakowski-Sudarshan conditions (\ref{LPM:LGKS}) may be more descriptively called completely positive rather than Markovian with reference to the property of the solution semigroup \cite{BaNaTh2008}.

Inspecting (\ref{LPM:CP}) immediately suggests the existence of a hierarchy of $k$-positive maps $\tilde{\Phi}^{(k)}(t,s)$, $k=1,\dots, d$ characterized by the conditions
\begin{align}
\tilde{\Phi}^{(k)}(t,s)[\tilde{\operatorname{O}}]:=(\mathrm{Id}_{k}\otimes\Phi(t,s))[\tilde{\operatorname{O}}]\,\geq\,0
\qquad \forall\,t\,\geq\,s
	\label{LPM:k_pos}
\end{align}
where now $\mathrm{Id}_{k}$ is the identity on $ \mathcal{M}_{k}(\mathbb{C})$ and the inequality holds for all positive matrices $\tilde{\operatorname{O}}$ in $\mathcal{M}_{kd}(\mathbb{C}) = \mathcal{M}_{k}(\mathbb{C}) \otimes \mathcal{M}_{d}(\mathbb{C})$. If we then denote by $\mathcal{P}_{k}$
the set of $k$-positive maps, it is rather straightforward to prove the chain of inclusions \cite{ChrD2022}
\begin{align}
	\mathcal{P}\,\equiv\,\mathcal{P}_{1}\supset \mathcal{P}_{2}\supset \dots\supset \mathcal{P}_{d}\,\equiv\,\mathcal{CP}
	\label{LPM:foliation}
\end{align}
We thus see that the class of $d$-positive maps, i.e., of completely positive maps ($ \mathcal{CP}$), is the smallest set in the above chain of inclusions. Unfortunately, for $k$-positive maps, there is no general counterpart of the operator-sum representation (\ref{LPM:povm}) that characterizes completely positive maps. Analogously, at the differential level, the conditions (\ref{LPM:LGKS}) have no general counterpart to specify generators of $k$-positive semigroups \cite{WoCi2008,ChrD2022}. Even if it can be argued that completely positive maps are the only ones adapted to describe the evolution of open quantum systems (see, however, the debate \cite{PecP1994,AliR1995,PecP1995} and the more recent \cite{DoLi2016}), completely positive dynamical maps can be solutions of a quantum master equation (\ref{LPM:qme}) whose generator does not satisfy the complete positivity conditions (\ref{LPM:LGKS}). The upshot is that quantum master equations that do not satisfy (\ref{LPM:LGKS}) must necessarily be considered. We refer to \cite{SmVa2010} for an example with a clear physical interpretation.
In addition, the study of $k$-positive maps for each $k$ is relevant to criteria for discriminating between entangled and separable states  (see \cite{ChKo2009} and the references therein). To summarize:
	\begin{enumerate}
		\item (\ref{LPM:gen}) equipped with regularity conditions on the time dependence is the generator of a semigroup of trace and self-adjointness preserving linear maps or, equivalently, of infinitely divisible trace and self-adjointness preserving linear maps;
		\item we call the semigroup trace preserving and positive if all its elements are linear maps that send the convex set of state operators into itself;
		\item the properties under tensor product composition with the identity map filter the set of positive linear maps into the flag of nested subsets (\ref{LPM:foliation}). Unfortunately, in general the resulting classification does not lead to an explicit representation;
		\item however, when (\ref{LPM:LGKS}) holds true the semigroup solution of a quantum master equation only consists of completely positive maps enjoying the property (\ref{LPM:CP}). This property is equivalent \cite{StiW1955,ChoM1975} to the Kraus positive operator sum (\ref{LPM:povm});
		\item in the time-autonomous case the semigroup enjoys the time translational invariance property (\ref{LPM:tti}): it thus reduces to a one parameter family of linear maps.
	\end{enumerate}
We conclude the section by recalling that a trace and self-adjointness preserving linear map $\Phi$ is positive if and only if it is a contraction \cite{KosA1972}
\begin{align}
\left\|\Phi(\operatorname{O})\right\|_{1}\,\leq\,\left\|\operatorname{O}\right\|_{1}
\qquad \forall\,\operatorname{O}=\operatorname{O}^{\dagger}\in\mathcal{M}_{d}(\mathbb{C})
	\label{LPM:contraction}
\end{align}
with respect to the trace norm (also known as Schatten $1$-norm)
\begin{align}
	\left\|\operatorname{O}\right\|_{1}:=\operatorname{Tr}\left(\sqrt{\operatorname{O}^{\dagger}\operatorname{O}}\right)
	\label{LPM:trace_norm}
\end{align}
The trace norm for self-adjoint operators coincides with the sum of the absolute values of all eigenvalues. Thus, positive linear maps are contractive and transform a convex compact set into itself. Again, these facts allow us to demonstrate that positive maps satisfy the hypotheses of the Perron-Frobenius theorem. The conclusion is that all the eigenvalues belong to the unit disk of the complex plane and are symmetric with respect to the real axis. Furthermore, there is at least one eigenvalue equal to one. The right eigenoperator associated with this eigenvalue is a state operator after appropriate normalization.

\section{Embedded classical master equation}
\label{sec:ECME}

As (\ref{LPM:qme}) is self-adjointness preserving, solutions that evolve from self-adjoint initial conditions at $t=0$
\begin{align}
	\bm{\rho}(0)=\sum_{i=1}^{d} p_{i}\,\bm{\varphi}_{i}\bm{\varphi}_{i}^{\dagger}
	\label{ECME:init}
\end{align}
at any later instant of time are amenable to the diagonal form
\begin{align}
	\bm{\rho}(t)=\sum_{i=1}^{d} \wp_{i}(t)\,\bm{\phi}_{i}(t)\bm{\phi}_{i}^{\dagger}(t)
	\label{ECME:sol}
\end{align}
where both the $\bm{\varphi}_{i}$'s and the $\bm{\phi}_{i}(t)$'s constitute a complete orthonormal basis of $\mathbb{C}^{d}$ and the $p_{i}$'s and $\wp_{i}$'s respectively denote the eigenvalues of $\bm{\rho}(0)$, $\bm{\rho}(t)$. 

As orthonormal bases are connected by unitary maps, there exists a family of unitaries $\mathsf{U}(t)$ such that
\begin{align}
	\bm{\phi}_{i}(t)=\mathsf{U}(t)\bm{\varphi}_{i}
	\label{ECME:unitary}
\end{align}
If we insert (\ref{ECME:unitary}) into (\ref{LPM:qme}) and denote by $\bm{p}$, $\bm{\wp}(t)$ the vectors whose entries are the eigenvalues of (\ref{ECME:init}), (\ref{ECME:sol}) respectively, a straightforward computation \cite{TeMa1992,WiTo1999} yields
	\begin{align}
 \begin{split}
 & \bm{\dot{\wp}}(t)=\mathsf{W}(t)\,\bm{\wp}(t)
 \\
 & \bm{\wp}(0)=\bm{p}\,.
 \end{split}
		\label{ECME:cme}
	\end{align}
At any time $t$	the $d\times d$ matrices $ \mathsf{W}(t)$ have the same structure as the generator of the classical master equation (\ref{LPM:cgen}): now they are column-diagonally dominated with elements
	\begin{align}
		\mathsf{W}_{i,j}(t)	=\left \langle\, \bm{\phi}_{i}(t)\,,\mathfrak{L}(t)[ \bm{\phi}_{j}(t)  \bm{\phi}_{j}^{\dagger}(t)]\, \bm{\phi}_{i}(t)\,\right\rangle	=
	\begin{cases}
\mathsf{R}_{i,j}(t)	& \forall\,i\neq j
		\\[0.3cm]
-\sum_{k\neq j}^{d} \mathsf{R}_{k,j}(t)	&\forall\, i=j
	\end{cases}
	\label{ECME:W}
	\end{align}
	where $\left \langle\,\cdot\,,\cdot\,\right\rangle$ is the standard inner product in $\mathbb{C}^{d}$ and the off-diagonal elements are the sum over the corresponding matrix elements of the decoherence operators weighted by the canonical couplings:
	\begin{align}
		\mathsf{R}_{i,j}(t)=\sum_{\ell=1}^{d^{2}-1}\mathscr{c}_{\ell}(t)\,\left |\left \langle\,\bm{\phi}_{i}(t)\,,\operatorname{L}_{\ell}(t)\bm{\phi}_{j}(t)\,\right\rangle \right |^{2}
		\label{ECME:rates}
	\end{align}
 Before continuing, some remarks are in order.
 \begin{itemize}
 \item In general, the {\textquotedblleft}spectral flow{\textquotedblright} equations (\ref{ECME:unitary}), (\ref{ECME:cme}) do not provide a constructive method for solving (\ref{LPM:qme}) but are only a tool to analyze the kinematic properties of individual solutions. More precisely, the matrix elements of (\ref{ECME:rates}) depend on the unitary evolution operator in (\ref{ECME:unitary}). The only general method to identify the latter requires first solving (\ref{LPM:qme}) for a given self-adjoint initial condition (\ref{ECME:init}). This also means that the families of matrices $\mathsf{U}(t)$ and $\mathsf{W}(t)$ depend on the choice of the initial conditions. In other words, the spectral flow equations play the same role for the kinematic analysis of quantum master equations as the current velocity formalism plays for the kinematic analysis of the Fokker-Planck equation \cite[\S~13]{NelE2001}.
 \item The derivation of (\ref{ECME:cme}) requires almost-everywhere differentiability of eigenvalues and eigenvectors of families of self-adjoint operators satisfying (\ref{LPM:qme}). Mathematically, any solution $\mathscr{x}(t)$ of a linear differential equation with respect to a real variable $t$ can be analytically continued to a unique holomorphic function of the independent variable:
	\begin{align}
	\overline{\mathscr{x}(z)}=\mathscr{x}(\overline{z})\qquad \forall\,z\in \mathbb{C} \quad\text{ where } \operatorname{Re}z=t\,.
		\label{ECME:holo_x}
	\end{align}
	As a consequence, we can analytically continue self-adjoint solutions of (\ref{LPM:qme}) to matrix valued holomorphic functions
	\begin{align}
		\bm{\rho}^{\dagger}(z)=\bm{\rho}(\bar{z})
		\label{ECME:holo}
	\end{align}
 	in a neighborhood of the real $t$-axis.	 This fact allows us to invoke Theorem 6.1 in \cite[\S~2.6.1]{KatT1982}, which guarantees that the eigenvalues and spectral projectors of a family of operators satisfying (\ref{ECME:holo}) are holomorphic functions and, as such, are continuous and differentiable. In addition \cite[\S~2.6.1]{KatT1982}, eigenvectors can also be chosen to be holomorphic functions, thus ensuring the validity of the construction used in the proof.
	\end{itemize}

\subsection{Kossakowski conditions}

The representation (\ref{ECME:sol}) yields the identity
\begin{align}
	\left\|\bm{\rho}(t)\right\|_{1}=\sum_{i=1}^{d}|\wp_{i}(t)|=\left\|\bm{\wp}(t)\right\|_{1}
	\label{ECME:contractive}
\end{align}
 holding for any solution of (\ref{LPM:qme}) with self-adjoint initial data. This means that (\ref{LPM:qme}) is contractive if and only if all classical master equations embedded in particular solutions are contractive.
 
It is readily clear that if the complete positivity conditions (\ref{LPM:LGKS}) hold true, then the off-diagonal entries (\ref{ECME:rates}) are positive functions of time. Consequently, (\ref{ECME:cme}) is a classical master equation and, as such, describes contractive dynamics. Conversely, if the entries (\ref{ECME:rates}) are always positive for any solution of (\ref{LPM:qme}) corresponding to self-adjoint initial data, then the semigroup generated by (\ref{LPM:qme}) is necessarily contractive and, therefore, positive \cite{ChrD2022}. In fact, requiring that the (\ref{ECME:rates}) are always positive is equivalent to the Kossakowski conditions \cite{KosA1972,GoKoSu1976} (see also \cite[\S~4.3]{RiHu2012}) asserting that a linear map of the form (\ref{LPM:gen}) is the generator of a trace-preserving contraction semigroup if and only if its matrix entries evaluated on an arbitrary orthonormal frame specify a classical generator. More explicitly, if we denote by $\big{\{}\operatorname{P}_{i}\big{\}}_{i=1}^{d}$ a complete set of orthonormal projectors on $\mathbb{C}^{d}$, that is,
\begin{align}
	\begin{split}
		&\sum_{i=1}^{d}\operatorname{P}_{i}=\operatorname{1}_{d} 
		\\
		& \operatorname{P}_{i}\operatorname{P}_{j}=\delta_{i,j}\operatorname{P}_{i} 
	\end{split}
&&	\qquad\forall\,i,j=1,\dots,d
	\label{ECME:proj}
\end{align}
then the Kossakowski conditions read
\begin{align}
	\begin{cases}
		\operatorname{Tr}\left(\operatorname{P}_{i}\mathfrak{L}(t)[\operatorname{P}_{j}]\right)\,\geq\,0 & i\neq j
		\\[0.3cm]
		\operatorname{Tr}\left(\operatorname{P}_{i}\mathfrak{L}(t)[\operatorname{P}_{i}]\right)=-\sum_{k\neq i=1}^{d}\operatorname{Tr}\left(\operatorname{P}_{k}\mathfrak{L}(t)[\operatorname{P}_{i}]\right)\,\leq\,0	& i= j
	\end{cases}
	\label{ECME:K}
\end{align}
To summarize: the identity (\ref{ECME:contractive}) allows us to explicitly relate the contractivity of the semigroup generated by (\ref{LPM:qme}) to that of solutions of the classical master equation (\ref{ECME:cme}) and, therefore, to derive Kossakowski's conditions.

\section{Conditions on the embedded classical master equation stemming from the positivity class of the semigroup}
\label{sec:conditions}

We now turn to derive some properties of (\ref{ECME:cme}) that follow if we assume that the semigroup, in addition to being trace preserving and positive, enjoys further properties. To this end, we recall that the Hilbert-Schmidt inner product of two matrices is defined by
\begin{align}
	\langle \operatorname{A}, \operatorname{B} \rangle := \operatorname{Tr}\!\left(\operatorname{A}^{\dagger}\operatorname{B}\right)
	\qquad \forall\, \operatorname{A}, \operatorname{B} \in \mathcal{M}_{d}(\mathbb{C})
	\label{conditions:HS}
\end{align}
Accordingly, the trace of $\mathfrak{L}(t)$ is
\begin{align}
	\operatorname{Tr}\mathfrak{L}(t) := \sum_{\mu = 1}^{d^2} \langle \operatorname{A}_{\mu}\,, \mathfrak{L}(t)[\operatorname{A}_{\mu}] \rangle
\end{align}
where $\{\operatorname{A}_{\mu}\}_{\mu=1}^{d^2}$ is a complete orthonormal basis of $\mathcal{M}_{d}(\mathbb{C})$. In particular, for any complete orthonormal basis $\{\bm{f}_{i}\}_{i=1}^{d}$ of $\mathbb{C}^{d}$, the collection of outer products $\{\bm{f}_{i}\bm{f}_{j}^{\dagger}\}_{i,j=1}^{d}$ constitutes a complete orthonormal basis of $\mathcal{M}_{d}(\mathbb{C})$.  Hence, we obtain
\begin{align}
	\operatorname{Tr}\mathfrak{L}(t)
	&=
	\sum_{i,j=1}^{d}\operatorname{Tr}\!\left((\bm{f}_{i}\bm{f}_{j}^{\dagger})^{\dagger}\mathfrak{L}(t)[\bm{f}_{i}\bm{f}_{j}^{\dagger}]\right) \notag\\
	&=
	\sum_{i,j=1}^{d}\left\langle \bm{f}_{i}, \mathfrak{L}(t)[\bm{f}_{i}\bm{f}_{j}^{\dagger}]\,\bm{f}_{j}\right\rangle 
	\label{conditions:trace0}
\end{align}
It is straightforward to verify that (\ref{conditions:trace0}) is also equal to the trace of the reshaped generator $\mathscr{L}(t)$ defined  by (\ref{LPM:reshaped}):
\begin{align}
	\operatorname{Tr}\mathfrak{L}(t)= \operatorname{Tr}\mathscr{L}(t)=-d\,\sum_{\ell=1}^{d^{2}-1}\mathscr{c}_{\ell}(t)\,\leq\,0
	\label{conditions:trace}
\end{align}
This result implies contractivity of the volume element for any $t\,\geq\,s$:
\begin{align}
&	\det \mathscr{F}(t,s)=\exp\left(\int_{s}^{t}\mathrm{d}u\,\operatorname{Tr}\mathscr{L}(u)\right)
	\label{conditions:volume}
\end{align}
where $\mathscr{F}(t,s)$ is the solution of (\ref{LPM:qds}). 

\subsection{Condition on the embedded classical master equation imposed by 2-positivity}

A linear map 
\begin{align}
	\bm{X}\colon\mathcal{M}_{d}(\mathbb{C})\mapsto\mathcal{M}_{d}(\mathbb{C})
	\label{conditions:lm}
\end{align}
is called $2$-positive if for any positive operator $\operatorname{O}$ belonging to $\mathcal{M}_{2d}(\mathbb{C})$ 
\begin{align}
	\operatorname{O}\,\geq\,0 \quad \Longrightarrow \quad(\mathrm{Id}_{2}\otimes\bm{X})(\operatorname{O})\,\geq\,0\,.
	\label{conditions:O}
\end{align}
Now, let us assume that (\ref{LPM:qme}) generates a $2$-positive semigroup. Then we claim that the classical master equation constructed out of any of the particular solutions is such that
\begin{align}
	0\,\geq\,\operatorname{Tr}\mathsf{W}(t)	\,\geq\, \frac{1}{d}\operatorname{Tr}\mathfrak{L}(t) 
		\label{conditions:2p}
\end{align}
An equivalent version of this claim was first proven in \cite{vEnChKiMG2025}. 

\subsubsection{Proof of the claim}

The action of a semigroup element close to the identity on a state operator $\bm{\rho}$ admits the explicit representation:
\begin{align}
	\Phi(t+\varepsilon,t)[\bm{\rho}]=\bm{\rho}+\varepsilon\,\mathfrak{L}(t)[\bm{\rho}]+O(\varepsilon^{2})
\end{align}
Let us denote by $\cab{i}{2} $, $i=1,2$ the elements of the canonical (or computational) basis of $\mathbb{C}^{2}$.  We can then construct a vector in $\mathbb{C}^{2d}$ of the form
\begin{align}
	\bm{u}=\frac{\cab{1}{2}\otimes \bm{\phi}_{i}(t)+\cab{2}{2}\otimes \bm{\phi}_{j}(t)}{\sqrt{2}}
\end{align}
where $\left\{ \bm{\phi}_{i}(t)\right\}_{i=1}^{d}$ is a complete orthonormal basis of $\mathbb{C}^{d}$ determined by the eigenvector flow equation (\ref{ECME:unitary}). The projector along $\bm{u}$ is clearly a positive matrix
\begin{align}
	\operatorname{P}_{u}=\bm{u}\bm{u}^{\dagger}
	=\frac{1}{2}\begin{bmatrix}
		 \bm{\phi}_{i}(t)  \bm{\phi}_{i}^{\dagger}(t) & \bm{\phi}_{i}(t)  \bm{\phi}_{j}^{\dagger}(t)	
		\\
		 \bm{\phi}_{j}(t)  \bm{\phi}_{i}^{\dagger}(t)	 
		&
		 \bm{\phi}_{j}(t)  \bm{\phi}_{j}^{\dagger}(t)
	\end{bmatrix}.
\end{align}
Then we find
\begin{align}
		(\mathrm{Id}_{2}\otimes\Phi(t+\varepsilon,t))[\operatorname{P}_{u}]&=
	\frac{1}{2}\begin{bmatrix}
		 \bm{\phi}_{i}(t)  \bm{\phi}_{i}^{\dagger}(t) & \bm{\phi}_{i}(t)  \bm{\phi}_{j}^{\dagger}(t)	
		\\
		 \bm{\phi}_{j}(t)  \bm{\phi}_{i}^{\dagger}(t)	 
		&
		 \bm{\phi}_{j}(t)  \bm{\phi}_{j}^{\dagger}(t)
	\end{bmatrix}
	\nonumber\\
&	+\frac{\varepsilon}{2}\begin{bmatrix}
		\mathfrak{L}(t)\big{[} \bm{\phi}_{i}(t)  \bm{\phi}_{i}^{\dagger}(t)\big{]}
		&
		\mathfrak{L}(t)\big{[} \bm{\phi}_{i}(t)  \bm{\phi}_{j}^{\dagger}(t)\big{]}
		\\
		\mathfrak{L}(t)\big{[} \bm{\phi}_{j}(t)  \bm{\phi}_{i}^{\dagger}(t)\big{]}
		&
		\mathfrak{L}(t)\big{[} \bm{\phi}_{j}(t)  \bm{\phi}_{j}^{\dagger}(t) \big{]}
	\end{bmatrix}
	+O(\varepsilon^{2})\,.
\end{align}
As the semigroup is $2$-positive by assumption, we must have
\begin{align}
	&	\left \langle\,\bm{v}\,,(\mathrm{Id}_{2}\otimes\Phi(t+\varepsilon,t))[\operatorname{P}_{u}]\,\bm{v}\,\right\rangle\,\geq\,0 && \forall\,\bm{v}\in\,\mathbb{C}^{2d}
\end{align}
In particular, we can choose 
\begin{align}
	\bm{v}=\frac{\cab{1}{2}\otimes \bm{\phi}_{i}(t)-\cab{2}{2}\otimes \bm{\phi}_{j}(t)}{\sqrt{2}}=\frac{1}{\sqrt{2}}\begin{bmatrix}
		 \bm{\phi}_{i}(t)	\\ - \bm{\phi}_{j}(t)
	\end{bmatrix}
\end{align}
so that
\begin{align}
	\left \langle\,\bm{u}\,,\bm{v}\,\right\rangle=0\,.
\end{align}
In such a case, after a straightforward calculation, we obtain
\begin{align}
	&0	\,\leq\,\lim_{\varepsilon\downarrow 0}\frac{\left \langle\,\bm{v}\,,(\mathrm{Id}_{2}\otimes\Phi(t+\varepsilon,t))[\operatorname{P}_{u}]\bm{v}\,\right\rangle}{\varepsilon}
	\nonumber\\
	&\hspace{0.5cm}=\frac{\left \langle\, \bm{\phi}_{i}(t)\,,\mathfrak{L}(t)\big{[} \bm{\phi}_{i}(t)  \bm{\phi}_{i}^{\dagger}(t)\big{]} \bm{\phi}_{i}(t)\,\right\rangle
		+\left \langle\, \bm{\phi}_{j}(t)\,,\mathfrak{L}(t)\big{[} \bm{\phi}_{j}(t)  \bm{\phi}_{j}^{\dagger}(t)\big{]} \bm{\phi}_{j}(t)\,\right\rangle}{4}
	\nonumber\\
	&\hspace{0.5cm}-\frac{\left \langle\, \bm{\phi}_{i}(t)\,,\mathfrak{L}(t)\big{[} \bm{\phi}_{i}(t)  \bm{\phi}_{j}^{\dagger}(t)\big{]} \bm{\phi}_{j}(t)\,\right\rangle+\left \langle\, \bm{\phi}_{j}(t)\,,\mathfrak{L}(t)\big{[} \bm{\phi}_{j}(t)  \bm{\phi}_{i}^{\dagger}(t)\big{]} \bm{\phi}_{i}(t)\,\right\rangle}{4}
\end{align}
Summing over $i,j=1,\dots,d$, the definitions (\ref{ECME:W}) and (\ref{conditions:trace}) yield the claim.

\subsection{Condition on the embedded classical master equation imposed by the Schwarz-class}
\label{sec:Schwarz}

It can be shown (see, e.g., the discussion in \cite{ChKiMu2024} and references therein) that the convex set of positive maps on $\mathcal{M}_{d}(\mathbb{C})$ yields a hierarchy of inclusions finer than (\ref{LPM:foliation}). For the purposes of the present work, we only note that between the convex sets of positive and $2$-positive maps lies the set $\mathcal{S}_{1}$ (also convex) of Schwarz maps:
\begin{align}
	\mathcal{P}\,\equiv\,\mathcal{P}_{1}\supset \mathcal{S}_{1}\supset\mathcal{P} _{2}
\end{align}
 We characterize Schwarz maps by considering their adjoint with respect to the Hilbert-Schmidt inner product (\ref{conditions:HS}). For an arbitrary linear map $\bm{X}$, we define its Hilbert-Schmidt adjoint $\bm{X}^{\ddagger}$ by requiring
\begin{align}
\langle \operatorname{A}\,,\bm{X}[\operatorname{B}]\rangle:=\operatorname{Tr}\left(\operatorname{A}^{\dagger}\bm{X}[\operatorname{B}]\right)=\operatorname{Tr}\left(\big{(}\bm{X}^{\ddagger}[\operatorname{A}]\big{)}^{\dagger}\operatorname{B}\right)=: \langle \bm{X}^{\ddagger}[\operatorname{A}]\,,\operatorname{B}\rangle\qquad \forall\,\operatorname{A},\operatorname{B}\in\mathcal{M}_{d}(\mathbb{C})
\end{align}
It is straightforward to verify that trace-preservation imposes the adjoint to be unital
\begin{align}
	\bm{X}^{\ddagger}[\operatorname{1}_{d}]=\operatorname{1}_{d}\,.
\end{align}
We say that a positive unital map is Schwarz class if 
\begin{align}
\bm{X}^{\ddagger}[\operatorname{A}^{\dagger}\operatorname{A}]\,\geq\,\bm{X}^{\ddagger}[\operatorname{A}^{\dagger}]\bm{X}^{\ddagger}[\operatorname{A}] \qquad \forall\,\operatorname{A}\,\in\,\mathcal{M}_{d}(\mathbb{C})
\end{align}
It is worth recalling here that Kadison's theorem asserts that the same inequality is satisfied by all positive maps \textit{if} the domain is restricted to normal operators \cite{KadR1952}. Conversely, $2$-positivity of unital maps implies the Schwarz property. 

Let us assume that a semigroup is Schwarz class. A straightforward calculation shows that the generator of such a semigroup must satisfy the dissipativity condition:
\begin{align}
\mathfrak{L}^{\ddagger}(t)[\operatorname{A}^{\dagger}\operatorname{A}]
	\,\geq\,
	\mathfrak{L}^{\ddagger}(t)[\operatorname{A}^{\dagger}]\operatorname{A}
	+
	\operatorname{A}^{\dagger}\mathfrak{L}^{\ddagger}(t)[\operatorname{A}]
	\qquad \forall\,t \quad \& \quad\forall\,\,\operatorname{A}\,\in\,\mathcal{M}_{d}(\mathbb{C})\,.
	\label{conditions:diss}
\end{align}
We refer to \cite{ChKiMu2024} for explicit examples. 

Now, let us suppose that (\ref{LPM:qme}) generates a semigroup in $\mathcal{S}_{1}$. Then we claim that the classical master equation constructed from any of the particular solutions is such that
\begin{align}
	0\,\geq\,	\operatorname{Tr}\mathsf{W}(t)\,\geq\,\frac{2}{d+1}\operatorname{Tr}\mathscr{L}(t)\,.
	\label{conditions:KS}
\end{align}
An equivalent version of this claim was first proven in \cite{ChvEnKiMG2025}. 

\subsubsection{Proof of the claim}

If we evaluate the dissipativity condition (\ref{conditions:diss}) on non-self-adjoint projectors constructed with
 eigenvectors of (\ref{ECME:sol}) at time $t$ 
\begin{align}
	\operatorname{P}_{i,j}(t)= \bm{\phi}_{i}(t)  \bm{\phi}_{j}^{\dagger}(t)
\end{align}
 we get
\begin{align}
	\mathfrak{L}^{\ddagger}(t)[\operatorname{P}_{j,j}(t)]
	\,\geq\,
	\operatorname{P}_{i,j}^{\dagger}(t)\mathfrak{L}^{\ddagger}(t)[\operatorname{P}_{i,j}(t)]
	+\mathfrak{L}^{\ddagger}(t)[\operatorname{P}_{i,j}^{\dagger}(t)]\operatorname{P}_{i,j}(t)\,.
\end{align}
From the left-hand side of the inequality we obtain:
\begin{align}
	\sum_{i=1}^{d}\sum_{j\neq i}^{d}
	\operatorname{Tr}\left(\operatorname{P}_{j,j}(t)
	\mathfrak{L}^{\ddagger}(t)\big{[}\operatorname{P}_{j,j}(t)\big{]}\right )=\sum_{i,j=1}^{d}(1-\delta_{i,j})
	\operatorname{Tr}\left(
	\mathfrak{L}(t)\big{[}\operatorname{P}_{j,j}(t)\big{]}\operatorname{P}_{j,j}(t)\right )
	=(d-1)\,\operatorname{Tr}\mathsf{W}(t)
\end{align}
The result must be larger than or equal to the corresponding expression for the right-hand side of (\ref{conditions:diss}) 
\begin{align}
	&	\sum_{i=1}^{d}\sum_{j\neq i}^{d}
	\operatorname{Tr}\left(\operatorname{P}_{j,j}(t)\big{(}\operatorname{P}_{i,j}^{\dagger}(t)\mathfrak{L}^{\ddagger}(t)[\operatorname{P}_{i,j}(t)]
	+\mathfrak{L}^{\ddagger}(t)[\operatorname{P}_{i,j}^{\dagger}(t)]\operatorname{P}_{i,j}(t)\big{)}
	\right )
	\nonumber\\
	&=	\sum_{i,j=1}^{d}(1-\delta_{i,j})
	\operatorname{Tr}\left(\operatorname{P}_{j,i}(t)\mathfrak{L}^{\ddagger}(t)[\operatorname{P}_{i,j}(t)]+
	\mathfrak{L}^{\ddagger}(t)[\operatorname{P}_{i,j}^{\dagger}(t)]\operatorname{P}_{i,j}(t)
	\right )
	=2\,\left(\operatorname{Tr}\mathscr{L}(t)-\operatorname{Tr}\mathsf{W}(t)\right)
\end{align}
The claim follows at once by comparing the two identities above.

\section{A distilled view on Lyapunov exponents}
\label{sec:Lyapunov}

We briefly recall the definition and the properties of Lyapunov exponents propaedeutic to the proof of our main result. In doing so, we mainly draw from \cite{AdrL1995} and the more recent \cite{BarL2017}.

Let $\mathsf{X}(t)$ be the Cauchy matrix solution of a time-non-autonomous differential system on a $d$-dimensional Euclidean vector space $\mathbb{E}$ with the initial condition given by the identity matrix on $\mathbb{E}$:
\begin{align}
	\begin{split}
		&\dot{\mathsf{X}}(t)=\mathsf{A}(t)\mathsf{X}(t)
		\\
		&\mathsf{X}(0)=\mathsf{1}_{d}
	\end{split}
	\label{Lyapunov:ode}
\end{align}
We assume that $\mathsf{A}(t)$ is a continuous and uniformly bounded square matrix for all $t\,\geq\,0$. Under these hypotheses,  if $\big{\{}\bm{g}_{i}\big{\}}_{i=1}^{d}$ is any basis of  $\mathbb{E}$, then the characteristic exponents 
\begin{align}
	&	\chi(\bm{g}_{i})=\limsup_{t\to\infty} \frac{1}{t}\ln \|\mathsf{X}(t)\bm{g}_{i} \|_{a} && i=1,\dots,d
\end{align}
are well defined quantities. The definition is independent of the choice of the norm $\left\|\cdot\right\|_{a}$ provided the norm is equivalent to the Euclidean norm which corresponds to $a$ equal to $2$.  We elaborate some more on norm equivalence in Appendix~\ref{app:norms}.

Lyapunov introduced the term {\textquotedblleft}\emph{normal basis}{\textquotedblright} for a basis $\big{\{}\bm{f}_{i}\big{\}}_{i=1}^{d}$ that minimizes the sum over all the characteristic exponents \cite{OseV1968,BeGaGiSt1980}:
\begin{align}
	\sum_{i=1}^{d}\chi(\bm{f}_{i})=\liminf_{t\to\infty} \frac{1}{t}\ln\det \mathsf{X}(t)=\liminf_{t\to\infty} \frac{1}{t}\int_{0}^{t}\mathrm{d}s\operatorname{Tr} \mathsf{A}(s)
\end{align}
Correspondingly, the exponents (\ref{Lyapunov:def}) computed on a normal basis are called Lyapunov exponents:
\begin{align}
	&	\lambda_{i}=\chi(\bm{f}_{i})&& i=1,\dots,d
		\label{Lyapunov:def}
\end{align} 
Geometrically, a normal basis filters $\mathbb{E}$ into a flag of  $n\,\leq\,d$ nested linear subspaces
\begin{align}
	\{0\}:=\mathbb{V}_{0}\subset	\mathbb{V}_{1}\subset \dots \subset \mathbb{V}_{n}=\mathbb{E}
\label{Lyapunov:flag}
\end{align}
each of which is spanned by a subset of the elements of the basis.  This means that each $\mathbb{V}_{i}$ corresponds to a distinct value of the Lyapunov exponents. Clearly, the flag has $d$ elements if the Lyapunov exponents are non-degenerate. In what follows, we label the exponents from one to $d$ to simplify the notation.

The QR decomposition of $\mathsf{X}(t)$, see e.g. \cite[Th~2.1.14]{HoJo2013}, gives a more computationally concrete meaning to Lyapunov exponents. 
Namely, it is always possible to uniquely factorize the  solution $ \mathsf{X}(t)$ of (\ref{Lyapunov:ode}) into a unitary matrix $ \mathsf{U}(t)$  and an upper triangular matrix $\mathsf{T}(t) $ having positive diagonal entries:
\begin{align}
	\mathsf{X}(t)=\mathsf{U}(t)\,\mathsf{T}(t)
\label{Lyapunov:QR}
\end{align}
Upon denoting by $ \mathsf{T}_{i,i}(t) $ the diagonal entries of $\mathsf{T}(t)$,  we arrive at
\begin{align}
&	\lambda_{i}=\limsup_{t\to\infty} \frac{1}{t}\ln \mathsf{T}_{i,i}(t)  &&\forall\, i=1,\dots,d
	\label{Lyapunov:QRexp}
\end{align}
Under the same hypotheses as above on $\mathsf{A}(t)$, {\textquotedblleft}\emph{dual Lyapunov exponents}{\textquotedblright} $\Gamma_{i}$ are also well defined for the Cauchy problem 
\begin{align}
	\begin{split}
		&\dot{\mathsf{Y}}(t)=-\mathsf{A}^{\dagger}(t)\mathsf{Y}(t)
		\\
		&\mathsf{Y}(0)=\mathsf{1}_{d}
	\end{split}
	\label{Lyapunov:aode}
\end{align}
Traditionally see e.g. \cite[\S~1.4]{ByViGrNe1966} or \cite[\S~I.1]{AdrL1995}, \eqref{Lyapunov:aode} is referred to as the {\textquotedblleft}\emph{adjoint equation}{\textquotedblright} to \eqref{Lyapunov:ode}. To avoid confusion with the concept of {\textquotedblleft}adjoint{\textquotedblright} used in section~\ref{sec:Schwarz},  we refer to (\ref{Lyapunov:aode}) as the Lyapunov {\textquotedblleft}\emph{dual system}{\textquotedblright}.
The Cauchy matrix solving \eqref{Lyapunov:aode} is the inverse of the adjoint of that solving  \eqref{Lyapunov:ode}.
\begin{align}
	\mathsf{Y}(t)=\mathsf{X}{}^{-1}{}^{\dagger}(t)
\end{align}
The important consequence is the preservation of the inner product
\begin{align}
	&	\left \langle\,\mathsf{Y}(t)\bm{w}\,,\mathsf{X}(t)\bm{v}\,\right\rangle=\left \langle\,\bm{w}\,,\bm{v}\,\right\rangle
	&& \forall \bm{v}\,,\bm{w}\in \mathbb{E}_{d}
	\label{Lyapunov:CS}
\end{align}
Upon applying the Cauchy-Schwarz inequality we get
\begin{align}
	&	\|\mathsf{Y}(t)\bm{w}\|_{2}^{2}\, \|\mathsf{X}(t)\bm{v}\|_{2}^{2}\ge \left|\left \langle\,\bm{w}\,,\bm{v}\,\right\rangle\right|^{2}
	&& \forall \bm{v}\,,\bm{w}\in \mathbb{E}_{d}
\end{align}
Hence, if we arrange the Lyapunov exponents of the primal dynamics (\ref{Lyapunov:ode}) in decreasing order and those of the dual dynamics (\ref{Lyapunov:aode}) in increasing order, and we choose $\bm{v}$ and $\bm{w}$ as the corresponding elements of normal bases spanning the two flags of nested subspaces, we generally obtain the inequalities:
\begin{align}
	&	\lambda_{i}+\Gamma_{i}\,\geq\,0 && i=1,\dots\,d
	\label{Lyapunov:Perron}
\end{align}
These inequalities raise the question of identifying the conditions under which
\begin{align}
	&	\lambda_{i}=-\Gamma_{i} && i=1,\dots,d
	\label{Lyapunov:Grobman}
\end{align}
Cardinal results in the theory of Lyapunov exponents due to Oseledets \cite{OseV1968,OseV2008} and Ragunathan \cite{RagM1979}, see also \cite{BeGaGiSt1980}, show that answering this question is equivalent to determining the conditions under which we can replace $\limsup$ with $\lim$ in the definition (\ref{Lyapunov:def}), or equivalently (\ref{Lyapunov:QRexp}), of the exponents. The answer identifies the class of Lyapunov {\textquotedblleft}\emph{regular}{\textquotedblright} systems, which enjoy the property that the limit 
\begin{align}
	\lim_{t\to\infty} \Big{(}\mathsf{X}^{\dagger}(t)\mathsf{X}(t)\Big{)}^{\frac{1}{2\,t}}=\mathsf{Q}
	\label{Lyapunov:regular}
\end{align}
exists and specifies a well-defined, self-adjoint, positive square matrix $\mathsf{Q}$.
We refer to \cite[Th~7.2.2]{BarL2017} for a proof of the equivalence of the characterization of regularity (\ref{Lyapunov:regular}) to the one used in \cite{OseV1968,BeGaGiSt1980} and to \cite[Th~4.3.1]{BarL2017} for a proof
that Lyapunov regularity is equivalent to (\ref{Lyapunov:Grobman}).

When the limit (\ref{Lyapunov:regular}) exists, the Lyapunov exponents coincide with the logarithms of the eigenvalues $\big{\{}\mathsf{q}_{i}\big{\}}_{i=1}^{d}$ of $\mathsf{Q}$:
\begin{align}
&	\lambda_{i}=-\Gamma_{i}=\ln \mathsf{q}_{i} && i=1,\dots,d
\end{align}
Furthermore, for a regular system, the sum of the Lyapunov exponents exactly specifies the growth rate of a volume form:
\begin{align}
	\sum_{i=1}^{d}\lambda_{i}=	\lim_{t\to\infty} \frac{1}{t}\int_{0}^{t}\mathrm{d}s\operatorname{Tr} \mathsf{A}(s)
\label{Lyapunov:vf}
\end{align}
The characterization of regularity (\ref{Lyapunov:regular}) is the basis of many algorithms used in the physics literature \cite{CeCeVu2009,PiPo2016} to compute Lyapunov exponents. Mathematically, (\ref{Lyapunov:regular}) is a stronger requirement than global existence and uniqueness of solutions (\ref{Lyapunov:ode}). 
Fortunately, it is possible to prove sufficient conditions for regularity only based on the properties of the matrix $\mathsf{A}(t)$.
Specifically, (\ref{Lyapunov:ode}) is Lyapunov regular if $\mathsf{A}(t)$, in addition to be continuous and bounded, enjoys one of the following properties:
\begin{enumerate}[style=unboxed,label={\upshape\bfseries C-\roman*}]
	\item \label{Lyapunov:C1} is time independent 
	\begin{align}
		&		\mathsf{A}(t)=\mathsf{A}(0)=\mathsf{A} && \forall\,t\,\geq\,0
	\end{align}
	as we can directly verify by diagonalization see e.g. \cite[Lemma~3.5.3]{AdrL1995};
	\item \label{Lyapunov:C2}  is periodic in time with period $T>0$:
	\begin{align}
		&	\mathsf{A}(t+T)=\mathsf{A}(t)  && \forall t\,\geq\,0
	\end{align}
	This claim can be verified by means of Floquet theorem see e.g. \cite[Lemma~3.5.3]{AdrL1995}.
	\item \label{Lyapunov:C3} $\mathsf{A}(t)$ is almost reducible to a time independent or periodic matrix. Almost reducible means that for any $\delta\,\geq\,0$ there exists a nonsingular and continuously differentiable matrix $\mathsf{L}_{\delta}(t)$ such that
	\begin{align}
		\mathsf{B}(t)+\mathsf{C}(t)=\mathsf{L}_{\delta}^{-1}(t)	\mathsf{A}(t)\mathsf{L}_{\delta}(t)-\mathsf{L}_{\delta}^{-1}(t)\dot{\phantom{L}}\hspace{-0.2cm}\mathsf{L}_{\delta}(t)
	\end{align}
	such that $\mathsf{B}(t) $ is either constant or periodic and
	\begin{align}
		\|\mathsf{C}(t)\|\,\leq\,\delta
	\end{align}
	The matrix norm may be any norm induced by a vector norm equivalent to the Euclidean norm. The proof of this last sufficient condition is more laborious \cite[Th~3.5.2]{AdrL1995}.
\end{enumerate}
 The sufficient conditions \ref{Lyapunov:C1}--\ref{Lyapunov:C3} encompass time-autonomous, periodic and almost periodic generators of quantum master equations. Generators typically, if not always, encountered in applications involving deterministic quantum master equations satisfy one of the sufficient conditions above.

More generally, Oseledets' \emph{multiplicative ergodic theorem} \cite{OseV1968,RagM1979,BeGaGiSt1980,OseV2008} guarantees the existence of the limit (\ref{Lyapunov:regular}) when, roughly speaking, the family of Cauchy matrices constitutes a statistically stationary, ergodic sequence of matrices. The theorem is  of central importance for non-autonomous linear systems obtained by linearization around a solution $\bm{x}(t)$ of a non-linear differential equation to study its stability versus infinitesimal change of the initial data. In this latter case the hypotheses of Oseledets' theorem are satisfied e.g. when $\bm{x}(t)$ is ergodic. Physics applications of Oseledets' theorem to non-linear dynamics are beyond the scopes of the present contribution and are extensively discussed in \cite{CeCeVu2009,PiPo2016}. We expect, however,  Oseledets' theorem to be relevant for studying ensembles of random quantum master equations \cite{DeLaTaChZy2019}.

\section{Main result}
\label{sec:main}

We are now in a position to prove results equivalent to those of \cite{ChvEnKiMG2025,vEnChKiMG2025}. Semigroups generated by (\ref{LPM:qme}) are always trace and self-adjointness preserving. In addition
\begin{proposition}
 If the elements of the semigroup are all contractive and Lyapunov regular, the Lyapunov exponents satisfy 
 \begin{align}
 	0\,=\,	\lambda_{0}\,\geq\,\lambda_{1}\,\geq\,\dots\,\geq\,\lambda_{d^{2}-1}
 \end{align}
 while, according to the definition of section~\ref{sec:Lyapunov}, the exponents of the dual system are
 \begin{align}
 	\Gamma_{d^{2}-1}=-\lambda_{d^{2}-1}\,\geq\, \dots \,\geq\,\Gamma_{1}=-\lambda_{1}\,\geq\,\Gamma_{0}=	-\lambda_{0}=0
 \end{align}
 		Furthermore,
\begin{enumerate}
	\item  if the semigroup is also $2$-positive, i.e., (\ref{conditions:2p}) holds, then the smallest Lyapunov exponent obeys the bound
	\begin{align}
		\lambda_{d^{2}-1}\,\geq\,\frac{1}{d}\sum_{i=1}^{d^{2}-1}\lambda_{i}
		\label{main:2p_ineq}
	\end{align}
		\item  if instead the semigroup is Schwarz class, i.e., (\ref{conditions:KS}) holds, then
		\begin{align}
			\lambda_{d^{2}-1}\,\geq\,\frac{2}{d+1}\sum_{i=1}^{d^{2}-1}\lambda_{i}
			\label{main:S_ineq}
		\end{align}
		\item finally, if the semigroup does not enjoy any additional property, then the lower bound on the smallest Lyapunov exponent is trivial:
		\begin{align}
			\lambda_{d^{2}-1}\,\geq\,\sum_{i=1}^{d^{2}-1}\lambda_{i}
			\label{main:c_ineq}
		\end{align}
\end{enumerate}
\end{proposition}
Before turning to the proof, some observations are in order.
\begin{itemize}
	\item Our hypotheses require contractivity for all the elements of the semigroup or, in other words, positive divisibility of the linear maps. 
	\item The dual exponents $\Gamma_{i}$'s specify the asymptotic decay rates of a contractive semigroup. The inequalities for the asymptotic decay rates read:
\begin{equation}
	\Gamma_{d^{2}-1} \leq \varkappa_{d} \sum_{\ell=1}^{d^2-1}\Gamma_\ell 
\end{equation}
where the universal prefactor $\varkappa_d$ is
\begin{equation}\label{mainres2}
	\varkappa_{d} = 
	\begin{cases} 
		\begin{array}{ll} 
		\dfrac {1}{d} & \ \ \mbox{for (\ref{main:2p_ineq})} 
			\\[0.4cm] 
			\dfrac {2}{d+1} & \ \ \mbox{for (\ref{main:S_ineq})} 
			\\[0.4cm]
			1 & \ \ \mbox{for (\ref{main:c_ineq})} 
			\end{array}
	\end{cases}
\end{equation}
\item If (\ref{LPM:qme}) is time-autonomous, that is,
\begin{align}
&	\mathfrak{L}(t)=\mathfrak{L} && \forall\,t
\end{align}
 then the Lyapunov exponents $\lambda_{i}$'s coincide with the real part of the eigenvalues of $\mathfrak{L}$:
 \begin{align}
 	\left\{\lambda_{0},\dots, \lambda_{d^{2}-1}\right\}=\operatorname{Re}\operatorname{Sp}\mathfrak{L}
 \end{align}
(see, e.g., \cite[\S~2.4.1]{PiPo2016}). This observation recovers the results of \cite{ChvEnKiMG2025,vEnChKiMG2025}.
\item If (\ref{LPM:qme}) is time periodic, i.e., there exists a $T>0$ such that
\begin{align}
&		\mathfrak{L}(t+T)=\mathfrak{L}(t) && \forall\,t
\end{align}
then the Lyapunov exponents $\lambda_{i}$'s coincide with the real part of the Floquet exponents of the monodromy matrix (see, e.g., \cite[\S~2.4.1]{PiPo2016}).
\end{itemize}

\subsection{Proof of the bounds}

We start by noticing that to prove the bound, it is sufficient to consider the asymptotic behavior of solutions of (\ref{LPM:qme}) evolving from self-adjoint initial operators. This is because we can always construct an orthonormal basis that consists only of self-adjoint elements to span $\mathcal{M}_{d}(\mathbb{C})$. 

The Lyapunov dual of the reshaped master equation (\ref{LPM:qds}) reads
	\begin{align}
		\begin{split}
			& 	\partial_{t}\mathscr{G}(t,s)=-\mathscr{L}^{\dagger}(t)\mathscr{G}(t,s)
			\\
			& 	\mathscr{G}(s,s)=\mathsf{1}_{d^{2}}
		\end{split}
		\qquad\forall \,t\,\geq\,s
	\end{align}
The corresponding equation governing the evolution for $t\,\geq\,0$ of operators  in $\mathcal{M}_{d}(\mathbb{C})$ from an initial condition at $t=0$, is
	\begin{align}
		\begin{split}
&	\bm{\dot{\gamma}}(t)=-\mathfrak{L}^{\ddagger}(t)[\bm{\gamma}(t)]
\\
& \bm{\gamma}(0)=\bm{\gamma}_{0}
	\end{split}
	\label{main:tame-}
		\end{align}
Under our hypotheses, (\ref{main:tame-}) generates a Lyapunov regular, unital, self-adjointness preserving and globally expansive semigroup. Unitality immediately implies
\begin{align}
	\Gamma_{0}=\lambda_{0}=0
\end{align}
Moreover, an arbitrary solution evolving from a self-adjoint initial condition $\bm{\gamma}_{0}$ is always amenable to the spectral decomposition
\begin{align}
	\bm{\gamma}(t)=\sum_{i=1}^{d} \mathscr{g}^{\smi{i}}(t)\,\bm{\chi}^{\smi{i}}(t)\bm{\chi}^{\smi{i}}{}^{\dagger}(t)\,.
\end{align}
where the $\mathscr{g}^{\smi{i}}(t) $'s are the real eigenvalues of $\bm{\gamma}(t)$, and the $\bm{\chi}^{\smi{i}}(t)$'s are the corresponding  orthonormal eigenvectors for any $t$.  Hence, proceeding as in section~\ref{sec:ECME}, we derive the eigenvalue flow equation
\begin{align}
	\begin{split}
	&	\bm{\dot{\mathscr{g}}}(t)=\,-\,\mathsf{W}^{\top}(t)\,\bm{\mathscr{g}}(t)
		\\
		&\bm{\mathscr{g}}(0)=\bm{\mathscr{g}}_{0}
	\end{split}
	\label{main:tacme-}
\end{align}
Here, $\mathsf{W}^{\top}(t) $ is the transpose of the generator of the classical master equation embedded in (\ref{LPM:qme}). 
Next, we write the solution as
\begin{align}
	\bm{\mathscr{g}}(t)=\mathsf{G}(t)\bm{\mathscr{g}}_{0}
\end{align}
using the time-ordered exponential
\begin{align}
	\mathsf{G}(t)=\mathcal{T}\exp \left(-\int_{0}^{t}\mathrm{d}s\,\mathsf{W}^{\top}(s)\right)\qquad t\,\geq\,0\,.
	\label{main:taflow}
\end{align}
Note that 
\begin{align}
	\mathsf{G}(t) = \mathsf{S}^{-1}{}^{\top}(t)
\end{align}
 where $\mathsf{S}(t)$ is the stochastic matrix solution of the Cauchy problem specified by (\ref{ECME:cme}). Clearly,  (\ref{main:taflow}) must also be expansive for any spectral flow equation. Now, we observe that the Frobenius norm induced by the Hilbert-Schmidt inner product corresponds to the standard $2$-norm for the population vector:
\begin{align}
	\left\|\bm{\gamma}(t)\right\|_{2}^{2}	=\operatorname{Tr}(\bm{\gamma}(t)^{2}) =\sum_{i=1}^{d}\mathscr{g}^{\smi{i}}{}^{2}(t)=\|\bm{\mathscr{g}}(t)\|_{2}^{2} 
\end{align}
 This identity puts us in a position to invoke Lyapunov regularity. We compute Lyapunov exponents of the adjoint dynamics using the Euclidean norm and obtain
\begin{align}
	\Gamma_{d^{2}-1}\,\equiv\,\lim_{t\to \infty}\frac{1}{t}\ln \|\bm{\gamma}(t)\|_{2} \,=\,\lim_{t\to \infty}\frac{1}{t}\ln \|\bm{\mathscr{g}}(t)\|_{2}
	\label{main:Lyapunov}
\end{align}
for trajectories evolving from $\bm{\mathscr{g}}_{0}$. As (\ref{main:tame-}) is expansive, all its Lyapunov exponents are necessarily positive. Therefore, we can use the expression of the volume form growth rate (\ref{Lyapunov:vf}) (applied to the adjoint equation;  see also Appendix~\ref{app:Lyapunov}) and the identity
\begin{align}
	\operatorname{Tr}\Big{(}-\mathsf{W}^{\top}(s)\Big{)}=\operatorname{Tr}\Big{(}-\mathsf{W}(s)\Big{)}
\end{align}
 to derive the chain of inequalities
\begin{align}
	\lim_{t\to \infty}\frac{1}{t}\ln \|\bm{\mathscr{g}}(t)\|_{2}\,\leq\,\lim_{t\to \infty}\frac{1}{t}\ln \det \mathsf{G}(t)=
	\lim_{t\to \infty}\frac{1}{t}\int_{0}^{t}\mathrm{d}s\,\operatorname{Tr}\Big{(}-\mathsf{W}(s)\Big{)}
\end{align}
This last inequality asserts that the largest Lyapunov exponent of an expansive flow is less than the sum over all the exponents. Thus, we arrive at the chain of inequalities:
\begin{align}
	\Gamma_{d^{2}-1}\,\equiv\,	-\lambda_{d^{2}-1}\,\leq\,\lim_{t\to \infty} \frac{1}{t}\int_{0}^{t}\mathrm{d}s\,\operatorname{Tr}\left(-\mathsf{W}(s)\right)\,\leq\,\lim_{t\to \infty}\frac{1}{t}
	\begin{cases}
		\dfrac{1}{d} \displaystyle\int_{0}^{t}\mathrm{d}s\,\operatorname{Tr}\left(-\mathfrak{L}(s)\right)	& \mbox{for (\ref{main:2p_ineq})}
		\\[0.4cm]
		\dfrac{2}{d+1} \displaystyle\int_{0}^{t}\mathrm{d}s\,\operatorname{Tr}\left(-\mathfrak{L}(s)\right)	&\mbox{for (\ref{main:S_ineq})}
	\end{cases} 
	\label{main:ub}
\end{align} 
Even if $\mathsf{W}$ depends on the initial data, the bound is in terms of the trace of the reshaped generator, which is clearly independent of any initial condition. We elaborate more on this in the remark~\ref{sec:remark} below.

The last step is to recall the general expression of the volume form (\ref{conditions:volume}), which, for Lyapunov regular
systems, gives:
\begin{align}
\sum_{i=1}^{d^{2}-1}\Gamma_{i}=\lim_{t\to \infty}\frac{1}{t}\int_{0}^{t}\mathrm{d}s\,\operatorname{Tr}\left(-\mathfrak{L}^{\dagger}(s)\right)	\,=\,	\lim_{t\to \infty}\frac{1}{t}\int_{0}^{t}\mathrm{d}s\,\operatorname{Tr}\left(-\mathfrak{L}(s)\right)	\,=\,-\sum_{i=1}^{d^{2}-1}\lambda_{i}
	\label{ECME:volume}
\end{align}
Comparing (\ref{ECME:volume}) with (\ref{main:ub}) yields the proof of (\ref{main:2p_ineq}) and (\ref{main:S_ineq}). 
The bound (\ref{main:c_ineq}) arises once again from the observation that for globally unstable dynamics the maximal Lyapunov exponent is bounded from above by the sum over all Lyapunov exponents.

\subsubsection{Remark on Lyapunov spectra of quantum master and spectral flow equations}
\label{sec:remark}

The Lyapunov spectrum of the semigroup defined by a quantum master equation consists of $d^{2}$ elements.
The image via this semigroup of any self-adjoint initial datum specifies a distinct spectral flow equation. 
The Lyapunov spectrum of each of the distinct spectral flow equations consists of only $d$ elements. This means that we cannot reconstruct the full Lyapunov spectrum of the semigroup from the spectral flow of an individual solution (\ref{ECME:sol}). Nevertheless, the maximal Lyapunov exponent of the semigroup cannot exceed that of a spectral flow associated to generic initial data.

\subsection{Proof using the Lozinskii-Dahlquist estimates}

Lozinskii-Dahlquist estimates \cite{LozS1958,DahG1959}, see also \cite[\S~IV.10.1]{ByViGrNe1966} or \cite[\S~IV.6]{AdrL1995}, provide an alternative way to prove our main result. Lozinskii-Dahlquist estimates yield upper and lower bounds on the growth rate of solutions of linear non-autonomous systems in terms of matrix norms induced by vector norms equivalent to the Euclidean. We write the estimates for a solution of (\ref{ECME:cme}):
\begin{align}
\exp \left(-\int _{0}^{t}\mathrm{d}s\,\mu_{a} \Big{(}-\mathsf{W}(s)\Big{)}\right)\leq \frac{\|\bm{\wp}(t)\|_{a}}{\|\bm{\wp}(0)\|_{a}}\leq \exp \left(\int _{0}^{t}\mathrm{d}s\,\mu_{a}  \Big{(}\mathsf{W}(s)\Big{)}\right)
		\label{main:LD}
\end{align}
where $\mu_{a}$ is the so-called {\textquotedblleft}\emph{logarithmic  norm}{\textquotedblright} which, however, is not a norm in the strict sense. For any matrix $\mathsf{M}$ in $\mathcal{M}_{d}(\mathbb{C})$ the logarithmic norm admits an explicit expression for the most commonly used matrix norms induced by vector norms equivalent to the Euclidean:
\begin{align}
	\mu_{a}(\mathsf{M}) = \lim_{\varepsilon \to 0^+} \frac{\|\mathsf{1}_{d}+ \varepsilon\,\mathsf{M}\|_{a} - 1}{\varepsilon}=\begin{cases}
		\max\limits_{1\,\leq\,j\,\leq\,d} \left( \mathsf{M}_{j,j} + \sum_{i \neq j} |\mathsf{M}_{i,j}| \right)	&  1-\mbox{norm}: a=1
		\\[0.3cm]		
		\lambda_{\mathrm{max}}\left ( \dfrac{\mathsf{M}+\mathsf{M}^{\top}}{2} \right)	&  \mbox{Euclidean-norm}: a=2
		\\[0.3cm]
		\max\limits_{1\,\leq\,i\,\leq\,d} \left( \mathsf{M}_{i,i} + \sum_{j \neq i} |\mathsf{M}_{i,j}| \right)	&  \mbox{Infinity-norm}: a=\infty
	\end{cases}
\end{align}
By  $\lambda_{\mathrm{max}}$ we denote the largest eigenvalue. If we evaluate the logarithmic norm for the classical generator $\mathsf{W}(t)$ in (\ref{ECME:cme}), we can take advantage of the fact that $\mathsf{W}(t)$ is a column diagonally dominant matrix. Furthermore, our assumption that $\mathsf{W}(t)$ is specified by a positive quantum semigroup by (\ref{ECME:contractive}) implies that the off-diagonal elements of $\mathsf{W}(t)$ are positive quantities. More explicitly,
\begin{align}
\mathsf{W}(t)=\begin{bmatrix}
	-\sum_{i\neq 1}^{d} \mathsf{R}_{i,1}(t) &	\mathsf{R}_{1,2}(t) & \dots \\
	\mathsf{R}_{2,1}(t) &	-\sum_{i\neq 2}^{d} \mathsf{R}_{i,2}(t) & \dots \\
	\vdots & \vdots & \ddots
	\end{bmatrix}\,,\quad \mathsf{R}_{i,j}(t) \,\geq\,0 \qquad \forall\,i\neq j=1,\dots,d\,.
\end{align}
Let us consider the lower bound in (\ref{main:LD}) in the case of the $\infty$-norm. Without loss of generality, we assume that the minimum is obtained for $i=1$. We get
\begin{align}
&	\ln\frac{\|\bm{\wp}(t)\|_{\scriptscriptstyle{\infty}}}{\|\bm{\wp}(0)\|_{\scriptscriptstyle{\infty}}}\,\geq\,\int_{0}^{t}\mathrm{d}s \left(\mathsf{W}_{1,1}(s)-\sum_{j\neq 1}^{d}|\mathsf{W}_{1,j}(s)|\right)\,=\,
\nonumber\\
&\int_{0}^{t}\mathrm{d}s \left(-\sum_{i\neq 1}\mathsf{R}_{i,1}(s)-\sum_{j\neq 1}^{d}\mathsf{R}_{1,j}(s)\right)\,\geq\,
\int_{0}^{t}\mathrm{d}s \left(-\sum_{j=1}^{d}\sum_{i\neq 1}\mathsf{R}_{i,j}(s)\right)=\int_{0}^{t}\mathrm{d}s\,\operatorname{Tr}\mathsf{W}(s)
\end{align}
We are thus in a position to apply (\ref{conditions:2p}), (\ref{conditions:KS}) to recover (\ref{main:2p_ineq}) and (\ref{main:S_ineq}).

Several observations are again in order.
\begin{itemize}
	\item Lozinskii-Dahlquist estimates hold independently of the hypothesis of Lyapunov regularity  \cite[\S~IV.10.1]{ByViGrNe1966}. Therefore, they allow us to conclude in general 
	\begin{align}
		\lambda_{d^{2}-1}\,\geq\,\limsup_{t\to \infty}\frac{1}{t}\begin{cases}
			\dfrac{1}{d} \displaystyle\int_{0}^{t}\mathrm{d}s\,\operatorname{Tr}\left(\mathfrak{L}(s)\right)	& \mbox{for (\ref{main:2p_ineq})}
			\\[0.4cm]
			\dfrac{2}{d+1} \displaystyle\int_{0}^{t}\mathrm{d}s\,\operatorname{Tr}\left(\mathfrak{L}(s)\right)	&\mbox{for (\ref{main:S_ineq})}
		\end{cases} 
	\end{align}
	\item As emphasized in \cite[\S~IV.10.1]{ByViGrNe1966}, exactly because of their generality Lozinskii-Dahlquist estimates are very rough for a generic system. This is consistent with the proof \cite{TsBl1997} that it is possible to construct co-cycles such that the computation of Lyapunov exponents can be mapped into an NP-hard problem. Nevertheless, in the examples of the ensuing section~\ref{sec:examples} we show that the bounds (\ref{main:2p_ineq}), (\ref{main:S_ineq}) are actually tight. This fact appears to us a consequence of the fact that already for stochastic matrices, i.e. for classical positive maps, the infinity norm indeed provides a tight but trivial bound which follows from probability conservation. The flag structure (\ref{LPM:foliation}) of quantum positive maps appears then to impose further spectral constraints on the exponents.  As we discuss in the conclusions, spectral properties of quantum positive maps call for further investigation.
	\item Finally, we observe that the more coarse-grained approach followed in \cite{MGKiCh2025} only yields 
\begin{align}
	\Gamma_{d^{2}-1}\,\leq\,\lim_{t\to \infty}\frac{1}{t}\int_{0}^{t}\mathrm{d}s\,\sum_{\ell=1}^{d^{2}-1}|c_{\ell}(s)|
\end{align}
for Lyapunov regular systems. When the conditions (\ref{LPM:LGKS}) are satisfied, the above expression readily recovers the result that holds for $2$-positive maps. More generally, we can rewrite the upper bound as
\begin{align}
	\Gamma_{d^{2}-1}\,\leq\,\frac{1}{d}\sum_{\ell=1}^{d^{2}-1}\Gamma_{\ell}+\lim_{t\to \infty}\frac{1}{t}\int_{0}^{t}\mathrm{d}s\,\sum_{\ell=1}^{d^{2}-1}\Big{(}|c_{\ell}(s)|-c_{\ell}(s)\Big{)}
\end{align}
The second term on the right-hand side is then a natural quantifier of the deviation from a completely positive evolution. Furthermore, it is readily related to the indicator
\begin{align}
	\mathcal{I}=\int_{0}^{\infty}\mathrm{d}s\,\sum_{\ell=1}^{d^{2}-1}\Big{(}|c_{\ell}(s)|-c_{\ell}(s)\Big{)}
\end{align}
introduced in \cite{RiHuPl2010} with the same purpose.
\end{itemize}

\section{Examples}
\label{sec:examples}

We discuss three examples. The first example shows the existence of completely positive semigroups that saturate the bound (\ref{main:2p_ineq}). The second example includes the construction of a positive semigroup that contains a completely positive element which saturates (\ref{main:c_ineq}). Finally, the third example illustrates the saturation of (\ref{main:c_ineq}) by a positive element of a positive semigroup.

\subsection{Completely positive semigroup saturating the bound (\ref{main:2p_ineq})}

Let us consider the unital, completely positive master equation on $\mathbb{C}^{2}$
\begin{align}
	\bm{\dot{\rho}}(t)=\sum_{i=1}^{3} \frac{r_{i}}{2}\left(\sigma_{i}\bm{\rho}(t)\sigma_{i}-\bm{\rho}(t)\right)
	\label{examples:cpme}
\end{align}
with
\begin{align}
&r_{i} \,\geq\,0&&\,i\,=1\,,2\,,3
\end{align}
and the $\sigma_{i}$'s denote as usual the Pauli matrices. We can equivalently rewrite (\ref{examples:cpme}) as a vector valued equation in $\mathbb{C}^{4}$:
\begin{align}
	\bm{\dot{\mathscr{r}}}(t)=\mathscr{L}
	\bm{\mathscr{r}}(t)
\end{align}
where, upon denoting by $\mathsf{1}_{2}$ the identity in $\mathcal{M}_{2}(\mathbb{C})$,
\begin{align}
	\mathscr{L}=\frac{1}{2}\begin{bmatrix}
		r_{3}\sigma_{3}-(r_{1}+r_{2}+r_{3})\operatorname{1}_{2} & r_{1} \sigma_{1}+\imath r_{2}\sigma_{2}	\\ r_{1} \sigma_{1}-\imath r_{2}\sigma_{2}	& -r_{3}\sigma_{3}-(r_{1}+r_{2}+r_{3})\operatorname{1}_{2}
	\end{bmatrix}
\end{align}
is the reshaped generator whose eigenvalues are
\begin{align}
&\lambda_{0}=\Gamma_{0}=0	
\nonumber\\
&	\lambda_{1}=-(r_{1}+r_{2}):=-\Gamma_{1}
\nonumber\\
 &\lambda_{2}=-(r_{1}+r_{3}):=-\Gamma_{2}
 \nonumber\\
&\lambda_{3}=-(r_{2}+r_{3}):=-\Gamma_{3}\,.
\end{align}
The upper bound (\ref{main:2p_ineq}) becomes
\begin{align}
	\sum_{i=1}^{3}\frac{\Gamma_{i}}{2}=r_{1}+r_{2}+r_{3}\,.
\end{align}
To saturate the bound we may choose
\begin{align}
	r_{1}=r_{2}=0
\end{align} 
which yields
\begin{align}
\Gamma_{0}=\Gamma_{1}=0 \qquad\Gamma_{2}=\Gamma_{3}=r_{3}\,.
\end{align}
We refer to \cite{ChKiKoSh2021} for a general method to construct master equations generating semigroups that saturate the bound in any dimension. Finally, we note that (\ref{examples:cpme}) can be interpreted as an error model in a quantum state subject to weak noise \cite[\S~1.4.5]{LiBr2009}. In particular, the case saturating the bound is commonly referred to as a purely dephasing channel.

\subsection{Saturation of the bound (\ref{main:c_ineq}) by a completely positive element of a non completely positive semigroup}

We now consider the quantum master equation
\begin{align}
	\bm{\dot{\rho}}(t)=\sum_{i=+,-,3}c^{\smi{i}}(t) \left(\sigma_{i}\bm{\rho}(t)\sigma_{i}^{\dagger}-\frac{\sigma_{i}^{\dagger}\sigma_{i}\bm{\rho}(t)+\bm{\rho}(t)\sigma_{i}^{\dagger}\sigma_{i}}{2}\right)
	\label{examples:cb}
\end{align}
with
\begin{align}
\sigma_{+}=\frac{\sigma_{1}+\imath\,\sigma_{2}}{2} \qquad\sigma_{-}=\frac{\sigma_{1}-\imath\,\sigma_{2}}{2} 
\end{align}
and coefficients continuous and bounded functions of the positive real semiaxis:
\begin{align}
&|c^{\smi{i} }(t)|\,<\,\infty &&i=+,-,3& \forall\,t\,\geq\,0
\end{align}
Under these hypotheses, (\ref{examples:cb}) always generates a semigroup of trace and self-adjointness preserving linear maps.
The equation (\ref{examples:cb}) is, however, not in canonical form because
\begin{align}
	\operatorname{Tr}\sigma_{+}^{\dagger}\sigma_{+}=\operatorname{Tr}\sigma_{-}\sigma_{+}=1
\end{align}
but
\begin{align}
	\operatorname{Tr}\sigma_{3}^{2}=2
\end{align}
If we represent the state operator in Bloch coordinates
\begin{align}
	\bm{\rho}(t)=\frac{\operatorname{1}_{2}+\sum_{i=1}^{3}\mathscr{x}^{\smi{i}}(t)\sigma_{i}}{2}
\end{align}
we find that the master equation is equivalent to the first-order system
\begin{align}
\bm{\dot{\mathscr{x}}}(t)=-\begin{bmatrix}
		0 & 0 & 0 & 0\\
	0 &c^{\smi{\perp}}(t) & 0 & 0\\
	0 & 0 &c^{\smi{\perp}}(t) & 0 \\
	c^{\smi{+}}(t)-c^{\smi{-}}(t) & 0 & 0 & c^{\smi{\parallel}}(t)
	\end{bmatrix}\bm{\mathscr{x}}(t)
	\quad\text{ where }\quad
	\bm{\mathscr{x}}(t)=\begin{bmatrix}
1	\\ \mathscr{x}_{1}(t)\\ \mathscr{x}_{2}(t)	\\ \mathscr{x}_{3}(t)
	\end{bmatrix}
	\label{examples:cbeq}
\end{align}
with
\begin{align}
	&c^{\smi{\perp}}(t)=\frac{c^{\smi{+}}(t)+c^{\smi{-}}(t)}{2}	+2\,c^{\smi{3}}(t)
	\label{examples:tr}
	\\
	&c^{\smi{\parallel}}(t)=c^{\smi{+}}(t)+c^{\smi{-}}(t)
	\label{examples:lo}
\end{align}
Since the generator is lower triangular, the generic element of the semigroup which evolves a solution of (\ref{examples:cbeq}) from an initial condition at an arbitrary earlier time $t_{0}\,\geq\,0$ to a later time $t$
\begin{align}
	\bm{\mathscr{x}}(t)=\mathscr{F}(t,t_{0})\bm{\mathscr{x}}(t_{0})
\end{align}
takes the form
\begin{align}
	\mathscr{F}(t,t_{0})=\begin{bmatrix}
	1 & 0 & 0 & 0	
	\\  
	0 & e^{-\int_{t_{0}}^{t}\mathrm{d}s\,c^{\smi{\perp}}(s)} & 0 & 0
	\\
	0 & 0 & e^{-\int_{t_{0}}^{t}\mathrm{d}s\,c^{\smi{\perp}}(s)} & 0 
	\\ 
	\int_{t_{0}}^{t}\mathrm{d}s\,	e^{-\int_{s}^{t}\mathrm{d}u\,c^{\smi{\parallel}}(u)}\Big{(}c^{\smi{+}}(s)-c^{\smi{-}}(s)\Big{)}	
	& 0 & 0 &e^{-\int_{t_{0}}^{t}\mathrm{d}s\,c^{\smi{\parallel}}(s)}
	\end{bmatrix}
	\label{examples:semigroup}
\end{align}
When (\ref{examples:tr}), (\ref{examples:lo}) are positive, we can refer to them as \emph{transversal} and \emph{longitudinal} rates, respectively. Positivity of  (\ref{examples:tr}), (\ref{examples:lo}) is a necessary condition for (\ref{examples:semigroup}) to be an element of a positive semigroup \cite{RuSzWe2002,WoCi2008}. This condition certainly holds if the $c^{\smi{i}}(t)$ for $i=+,-,3$ are positive functions. This, in turn, implies that the canonical couplings are also positive, i.e. satisfy (\ref{LPM:LGKS}). Therefore, (\ref{examples:cb}) generates a completely positive semigroup. In such a case, the following inequality always holds:
\begin{align}
	2\,c^{\smi{\perp}}(t)\,\geq\,c^{\smi{\parallel}}(t)\,\geq\,0
\end{align}
This inequality generalizes to the time-non-autonomous case a well known inequality satisfied by the relaxation times, i.e. the inverse of the rates, of a qubit state operator solution of a completely positive and time independent master equation, see e.g. \cite{PecP1994} or \cite[\S~8]{NiCh2010}. In addition, if the canonical couplings satisfy (\ref{LPM:LGKS}), the bound (\ref{main:2p_ineq}) becomes
\begin{align}
\lim_{t\to\infty}&\frac{1}{t}\int_{0}^{t}\mathrm{d}s \Big{(}c^{\smi{+}}(s)+c^{\smi{-}}(s)+2\,c^{\smi{3}}(s)\Big{)}\,\geq\,
\nonumber\\
&\operatorname{minimum}\left(\lim_{t\to\infty}\frac{1}{t}\int_{0}^{t}\mathrm{d}s\,c^{\smi{\perp}}(s) \,, \lim_{t\to\infty}\frac{1}{t}\int_{0}^{t}\mathrm{d}s\,c^{\smi{\parallel}}(s) \right)
\end{align} 
which is readily verified.

Let us now turn to the case in which not all canonical couplings are positive functions. In particular, we assume
\begin{align}
	c^{\smi{\pm}}(t)=1 
	\label{examples:c+-}
\end{align}
and
\begin{align}
	c^{\smi{3}}(t)=-\frac{1}{2}\tanh t \,.
	\label{examples:c3}
\end{align}
With this choice of the canonical couplings, the master equation does not generate a completely positive semigroup. Specifically, while $\Phi(t,0)$ is completely positive for $t > 0$, the maps $\Phi(t,s)$ fail to be completely positive for $t> s > 0$. We can, however, readily verify from (\ref{examples:semigroup}) that when we insert (\ref{examples:c+-}) and (\ref{examples:c3}), the semigroup consists of trace preserving diagonal matrices which map the Bloch sphere into the Bloch ball. This fact ensures the preservation of the set of state operators and is, therefore, equivalent to proving that the semigroup is positive \cite{RuSzWe2002}. In other words, all the maps $\Phi(t,s)$ are positive for any $t> s > 0$. For an alternative criterion to prove that a semigroup is positive, we refer to \cite[Th~25]{WoCi2008}.

Let's now turn to the analysis of the exponents. Trace-preservation always forces one Lyapunov exponent to vanish. The non-trivial asymptotic decay rates are
\begin{align}
\Gamma_{1}=	\Gamma_{2}=	\lim_{t\to \infty}\frac{1}{t}\int_{0}^{t}\mathrm{d}s\,c^{\smi{\perp}}(s)=1-\lim_{t\to \infty}\frac{\ln \cosh t}{t}=0
\end{align}
and
\begin{align}
\Gamma_{3}=\lim_{t\to \infty}\frac{1}{t}\int_{0}^{t}\mathrm{d}s\,c^{\smi{\parallel}}(s)=2
\end{align}
Together, the above results imply saturation of the bound (\ref{main:c_ineq}), a witness of a positive semigroup:
\begin{align}
2=	\Gamma_{3}=\sum_{\iota=1}^{3}\Gamma_{i}
\end{align}
Finally, we verify that the linear map $\Phi(t,0)$ is completely positive, as mentioned above.
Namely, solutions of (\ref{examples:cb}) evolving from initial conditions imposed at $t=0$
\begin{align}
	\bm{\rho}(t)=\frac{\operatorname{1}_{2}+e^{-t}\,\cosh t\, \big{(}\mathscr{x}_{0}^{\smi{1}}\,\sigma_{1}+\mathscr{x}_{0}^{\smi{2}}\,\sigma_{2}\big{)} +e^{-2t}\,\mathscr{x}_{0}^{\smi{3}}\,\sigma_{3} }{2}
\end{align}
specify the dynamical map
\begin{align}
&\bm{\mathscr{x}}(t)=\mathsf{F}(t)\bm{\mathscr{x}}(0) && \& & \mathsf{F}(t)=\begin{bmatrix}
	1 & 0 & 0 & 0	\\ 
	0 &e^{-t} \cosh t & 0 &0 \\
	0 & 0 &e^{-t}\cosh t &0 \\
	0 & 0 & 0 &e^{-2t}
\end{bmatrix}
\end{align}
To prove complete positivity, it is sufficient to verify that the spectrum of the Choi matrix  is positive \cite{ChoM1975} or, e.g., \cite[\S~10]{BeZy2006}. As we computed the dynamical map $\mathsf{F}(t)$ in the orthonormal basis of the matrix space $\mathcal{M}_{2}(\mathbb{C})$ specified by Pauli matrices, the Choi matrix takes the form (see, e.g., \cite[\S~B]{LiSeKoOnCu2025})
\begin{align}
	\mathsf{C}(t)=\sum_{i,j=1}^{4} \frac{\mathsf{F}_{j,i}(t)}{2}\overline{\sigma_{i-1}}\otimes\sigma_{j-1}
\end{align}
where $\sigma_{0}=\mathsf{1}_{2}$. We thus arrive at
\begin{align}
	\mathsf{C}(t)=	\begin{bmatrix}
		\frac{1+e^{-2t}}{2} & 0 &0 & \frac{1+e^{-2t}}{2} 	\\ 0 &\frac{1-e^{-2t}}{2}& 0 &0 \\ 0 & 0 & \frac{1-e^{-2t}}{2}& 0 \\ 	\frac{1+e^{-2t}}{2} & 0 & 0 & \frac{1+e^{-2t}}{2}
	\end{bmatrix}
\end{align}
whose spectrum
\begin{align}
	\operatorname{Sp}\mathsf{C}(t)=\left\{ 0\,,\frac{1-e^{-2t}}{2}\,,\frac{1-e^{-2t}}{2}\,,1+e^{-2t} \right\}
\end{align}
is positive, thus signaling complete positivity of the evolution from $t=0$. 

We thus verify that an element of a non completely positive semigroup can be completely positive. At the same time, the bound on Lyapunov exponents carries information about the semigroup and not about particular solutions of the master equation that generates it.

\subsection{Saturation of the bound (\ref{main:c_ineq}) by a positive element of a positive semigroup}

We consider a quantum master equation of the non-canonical form (\ref{examples:cb}) with $c^{\smi{\pm}}$ as in the previous example, but now
\begin{align}
	c^{\smi{3}}= -\frac{1}{2}\,.
\end{align}
The master equation is time-autonomous, hence any solution only depends on the time elapsed from the assignment of initial data:
\begin{align}
		\bm{\rho}(t)=\frac{\operatorname{1}_{2}+ \mathscr{x}_{0}^{\smi{1}}\,\sigma_{1}+\mathscr{x}_{0}^{\smi{2}}\,\sigma_{2} +e^{-2t}\,\mathscr{x}_{0}^{\smi{3}}\,\sigma_{3} }{2}
\end{align} 
which defines the dynamical map and Choi matrices
\begin{align}
	& \mathsf{F}(t)=\begin{bmatrix}
		1 & 0 & 0 & 0\phantom{e^{-}}	\\ 
		0 &1 & 0 &0 \phantom{e^{-}}\\
		0 & 0 &1 &0 \phantom{e^{-}}\\
		0 & 0 & 0 &e^{-2t}
		\end{bmatrix}
		&& \& & \mathsf{C}(t)=\begin{bmatrix}
			\frac{1+e^{-2t}}{2} & 0 &0 & 1	\\ 0 &\frac{1-e^{-2t}}{2}& 0 &0 \\ 0 & 0 & \frac{1-e^{-2t}}{2}& 0 \\ 	1 & 0 & 0 & \frac{1+e^{-2t}}{2}
		\end{bmatrix}
\end{align}
The dynamical map is contractive with rates
\begin{align}
&	\Gamma_{0}=\Gamma_{1}=\Gamma_{2}=0
	\nonumber\\
&	\Gamma_{3}=2
\end{align}
which readily saturate the bound (\ref{main:c_ineq}). The Choi matrix has one negative eigenvalue:
\begin{align}
		\operatorname{Sp}\operatorname{C}(t)=\left\{-\frac{1-e^{-2 t}}{2} ,\frac{1-e^{-2 t}}{2},\frac{1-e^{-2 t}}{2},\frac{3+e^{-2 t}}{2} \right\}
\end{align} 
A general result of \cite{JoLoPu2019} states that the Choi matrix of a $ k$-positive map on $\mathcal{M}_{d}(\mathbb{C})$ has at most $(d-k)^2$ negative eigenvalues. Indeed, the presence of one negative eigenvalue in the example shows that the solution can only be $1$-positive.

\section{Conclusions and outlook}

Linear maps preserving state operators are commonly referred to as trace preserving and positive. In comparison to the classical case, when positive maps are matrices sending the simplex of probability vectors into itself,  positive maps on state operators are considerably less understood. Most of the results available refer to the subclass of completely positive maps. These latter ones enjoy the appealing property of admitting an explicit algebraic representation, the Kraus positive operator sum (\ref{LPM:povm}).
This representation is instrumental to the proof, originally due to Lindblad \cite{LinG1976} and Gorini Kossakowski and Sudarshan  \cite{GoKoSu1976}, of the one-to-one correspondence between semigroups of trace preserving completely positive maps and explicit conditions, specifically (\ref{LPM:LGKS}), that fully characterize the canonical form of the infinitesimal generators. A recurrently debated question is whether the sub-class of completely positive maps should be identified with physically realizable evolution laws \cite{PecP1994,AliR1995,PecP1995,DoLi2016}. Without entering this debate, here we wish to emphasize that completely positive maps may be constructed as particular solutions of quantum master equations that do not enjoy the conditions (\ref{LPM:LGKS}) \cite{HaCrLiAn2014}. In other words, we have in mind the following scenario. We prepare the system at an initial time $t_0$, and we are interested in the evolution only for later values of the time variable $t$. The evolution is described by a family of trace-preserving and completely positive maps $\Phi(t,t_0)$. If these maps solve a master equation whose canonical couplings $c_\ell(t)$ are positive for any $t\geq t_0$, then not only $\Phi(t,t_0)$ but all elements $\Phi(t,s)$ with $t \geq s \geq t_0$ of the semigroup generated by the master equation are trace-preserving and completely positive. Therefore, the evolution is completely positive infinitely divisible \cite{HaCrLiAn2014,WoCi2008,ChrD2022}. For non completely positive infinitely divisible dynamics, the map $\Phi(t,t_0)$ is still completely positive but at least some semigroup elements $\Phi(t,s)$ fail to be completely positive. The characterization of the corresponding generators is notoriously difficult. Clearly, some couplings $c_\ell(t)$ have to be, at least temporally, negative for $t \geq t_0$, as in our second example.   
These considerations motivate the relevance of investigating the properties of semigroups generated by quantum master equations which do not satisfy the conditions (\ref{LPM:LGKS}). Additionally, they highlight the inverse problem of identifying which systems of first-order ordinary differential equations are amenable to the form of quantum master equations \cite{KaGuLi2023}. Related to this, our results specify necessary conditions for a semigroup to satisfy a completely positive master equation.
  
More generally, Perron-Frobenius theorem allows us to conclude that a quantum master equation generating a positive semigroup is equivalent to a dynamical system of the form (\ref{LPM:qds}) specified by a Hurwitz-stable matrix $\mathscr{L}$. The matrix must also satisfy additional constraints imposed by trace- and self-adjointness preservation. The bound that we present in our work provides a straightforward test in terms of experimentally measurable quantities to discriminate between semigroups of positive maps. We regard our result as a very first step towards addressing the question concerning to which extent the positivity class of the generated semigroup reflects on the spectral properties of the generator \cite{ChKo2009}. Tomographic reconstruction of state operators is also subject to state preparation and measurement errors. This consideration poses additional questions concerning the robustness of any spectral characterization (see, e.g., \cite{vEnF2024} for questions arising from eigenvalue degeneration).  Research addressing these topics is at the crossroads between different disciplines for which the study of stability of dynamical systems is relevant. In the literature there exist distinct notions of stability and they lead to the identification of matrix classes based on spectral properties \cite{KusO2019}. Do these notions provide insight on the positivity property of semigroups generated by quantum master equations? 

In the present work we illustrate a first step toward addressing the above research questions. We hope that our result may attract the attention on quantum master equations also from scholars whose interest in dynamical systems is motivated by applications to other disciplines.

\section{Acknowledgments}

The authors are grateful to Angelo Vulpiani for reading and commenting on the manuscript.
PMG acknowledges support from the CoE in Randomness and Structures (FiRST) of the Research Council of Finland (funding decision number: 346305).
GK was supported by JSPS KAKENHI Grant No. 24K06873.
FvE is funded by the \textit{Deutsche Forschungsgemeinschaft} (DFG, German Research Foundation) -- project number 384846402, and supported by the Einstein Foundation (Einstein Research Unit on Quantum Devices) and the MATH+ Cluster of Excellence. DC was supported by the Polish National Science Center under Project No. 2024/55/B/ST2/01781.

\appendix

\section{Vector norms equivalent to the Euclidean and their induced matrix norms}
\label{app:norms}

 \begin{definition}
 	The norm $\left\|\bm{v}\right\|_{a}$ of any $\bm{v}\in \mathbb{C}^{d}$ is \textbf{equivalent} to the Euclidean norm if there exist universal constants $0\,<c\,\leq\,C\,<\infty$ such that
 	\begin{align}
 		c\,\left\|\bm{v}\right\|_{a}\,\leq\,\left\|\bm{v}\right\|_{2}\,\leq\,C\,\left\|\bm{v}\right\|_{a}
 		\label{norms:equivalence}
 	\end{align}
 \end{definition} 
 Let $\big{\{}\bm{f}_{i}\big{\}}_{i=1}^{d}$ be a normal basis for the system (\ref{Lyapunov:ode}) and $\mathsf{X}(t)$  solution of (\ref{Lyapunov:ode}). The use of the inequalities (\ref{norms:equivalence}) readily show that Lyapunov exponents do not depend on the choice of the norm as long as it is equivalent to the Euclidean norm:
\begin{align}
\chi(\bm{f}_{i})=\limsup_{t\to\infty} \frac{1}{t}\ln \left\|\mathsf{X}(t)\bm{f}_{i}\right\|_{2}\,=
\limsup_{t\to\infty} \frac{1}{t}\ln \left\|\mathsf{X}(t)\bm{f}_{i}\right\|_{a}
\label{norms:Lyapunov}
\end{align}
Finding bounds on Lyapunov exponents raises the question of determining the norm of the Cauchy matrix $\mathsf{X}(t)$ solving (\ref{Lyapunov:ode}). Three matrix norms are particularly useful (see \cite[\S~I.2 \& \S~IV.6]{AdrL1995} for details). In listing these norms, we denote by $\mathsf{M}$ a $d\times d$ complex matrix.
\begin{enumerate}
	\item The matrix $1-$norm
	\begin{align}
		\left\|\mathsf{M}\right\|_{1}=\max_{1\,\leq\,i\,\leq\,\,d}\sum_{j=1}^{d}\left |\mathsf{M}_{j,i}\right |
		\label{norms:1n}
	\end{align}
	that is induced by the {\textquotedblleft}taxicab{\textquotedblright} norm on $\bm{v}\in \mathbb{C}^{d}$:
	\begin{align}
		\left\|\bm{v}\right\|_{1}=\sum_{i=1}^{d}|\bm{v}_{i}|
	\end{align}
	\item The matrix $2-$norm
	\begin{align}
		\left\|\mathsf{M}\right\|_{2}=\max_{\bm{v}\colon \left\|\bm{v}\right\|=1}\left\|\mathsf{M}\bm{v}\right\|_{2}
		\label{norms:2n}
	\end{align}
	with the vector norm on the right-hand side being the one induced by the standard inner product 
	\begin{align}
		\left\|\mathsf{M}\bm{v}\right\|_{2}^{2}=\left \langle\,\mathsf{M}\bm{v}\,,\mathsf{M}\bm{v}\,\right\rangle
	\end{align}
	Clearly, the $2$-norm coincides with the largest singular value of $ \mathsf{M}$, that is, 
	\begin{align}
		\left\|\mathsf{M}\right\|_{2}^{2}=\max_{1\,\leq\,i\,\leq\,d}\left\{ \mathscr{a}_{i}^{2}\hspace{0.2cm}\big{|} \hspace{0.2cm} \mathscr{a}_{i}^{2}\in\operatorname{Sp} \mathsf{M}^{\dagger}\mathsf{M}\right\}
	\end{align}
	\item The matrix {\textquotedblleft}infinity{\textquotedblright} norm:
	\begin{align}
		\left\|\mathsf{M}\right\|_{\mathfrak{\infty}}=\max_{1\,\leq\,i\,\leq\,d}\sum_{j=1}^{d}\left |\mathsf{M}_{i,j}\right |
		\label{norms:infn}
	\end{align}
	that is induced by the infinity norm on $\bm{v}\in \mathbb{C}^{d}$:
	\begin{align}
		\left\|\bm{v}\right\|_{\mathfrak{\infty}}=\max_{1\,\leq\,i\,\leq\,d}|\bm{v}_{i}|
	\end{align}
\end{enumerate}

\section{Computation of the spectrum Lyapunov exponents}
\label{app:Lyapunov}

Throughout this appendix we assume Lyapunov regularity. 

As discussed in section~\ref{sec:Lyapunov} of the main text, distinct Lyapunov exponents filter a a $d$-dimensional vector space $\mathbb{E}$ into the flag structure (\ref{Lyapunov:flag}). This means that for a generic element $\bm{x}$ of $\mathbb{E}$, we get
\begin{align}
	\lambda_{\mathrm{max}}=\lim_{t\to \infty} \frac{1}{t} \ln \|\mathsf{X}(t)\bm{x}\|_{2}
\end{align}
where $X(t)$ solves (\ref{Lyapunov:ode}). The limit is independent of $\bm{x}$ up to a set of measure zero. Namely, a generic initial condition $\bm{x}$ has a non-vanishing projection on the eigenvector $\bm{q}_{\mathrm{max}}$ corresponding to the largest eigenvalue of the matrix $\mathsf{Q}$ defined by the limit (\ref{Lyapunov:regular}). Let's assume
for simplicity sake that the spectrum of $\mathsf{Q}$ is non-degenerate. Applying the Cauchy matrix $\mathsf{X}(t)$ to a generic element of the orthogonal complement of $\bm{q}_{\mathrm{max}}$ yields the sub-dominant Lyapunov exponent. 
The orthogonal complement has dimension $d-1$ and has, therefore, vanishing Lebesgue measure. Upon iterating this idealized procedure we can compute all the Lyapunov exponents. In practice, to achieve the same goal we pick a generic complete orthonormal system $\big{\{}\bm{f}_{i}\big{\}}_{i=1}^{d}$ in $\mathbb{E}$ and upon assuming
\begin{align}
	\lambda_{\mathrm{max}}=\lambda_{1}\,>\,\lambda_{2}\,>\,\dots\,>\,\lambda_{d}
\end{align}
we compute for $j\,\leq\,d$
\begin{align}
\begin{split}
	& \lambda_{1}=\lim_{t\to \infty} \frac{1}{2t} \ln \left \langle\,\mathsf{X}(t)\bm{f}_{1}\,,\mathsf{X}(t)\bm{f}_{1}\,\right\rangle
	\\
	& \lambda_{1}+\lambda_{2}=\lim_{t\to \infty} \frac{1}{2 \,t} \ln \det \begin{bmatrix}
		\left \langle\,\mathsf{X}(t)\bm{f}_{1}\,,\mathsf{X}(t)\bm{f}_{1}\,\right\rangle	& 
		\left \langle\,\mathsf{X}(t)\bm{f}_{1}\,,\mathsf{X}(t)\bm{f}_{2}\,\right\rangle	
		\\ 
		\left \langle\,\mathsf{X}(t)\bm{f}_{2}\,,\mathsf{X}(t)\bm{f}_{1}\,\right\rangle	& 
		\left \langle\,\mathsf{X}(t)\bm{f}_{2}\,,\mathsf{X}(t)\bm{f}_{2}\,\right\rangle	
	\end{bmatrix}
	\\
	\vdots 
	\\
	& \lambda_{1}+\dots+\lambda_{j}=\lim_{t\to \infty} \frac{1}{2 \,t} \ln \det \begin{bmatrix}
		\left \langle\,\mathsf{X}(t)\bm{f}_{1}\,,\mathsf{X}(t)\bm{f}_{1}\,\right\rangle	& \dots &
		\left \langle\,\mathsf{X}(t)\bm{f}_{1}\,,\mathsf{X}(t)\bm{f}_{j}\,\right\rangle	
		\\
		\vdots & \ddots & \vdots 
		\\
		\left \langle\,\mathsf{X}(t)\bm{f}_{j}\,,\mathsf{X}(t)\bm{f}_{1}\,\right\rangle	& \dots &
		\left \langle\,\mathsf{X}(t)\bm{f}_{j}\,,\mathsf{X}(t)\bm{f}_{j}\,\right\rangle	
	\end{bmatrix}
\\
\vdots
\end{split}
\end{align}
For $j=d$ we obtain the growth rate of a volume form in $\mathbb{E}$. Namely, the identity
\begin{align}
	\ln\det \mathsf{M}=\operatorname{Tr}\ln \mathsf{M}
\end{align}
, which holds for any matrix $\mathsf{M}$, yields the equation
\begin{align}
	\frac{\mathrm{d}}{\mathrm{d} t}\ln \det \mathsf{X}(t)=\operatorname{Tr}\mathsf{A}(t)
\end{align}
Thus, it is generically possible to derive an explicit expression for the sum over all Lyapunov exponents using the solution of the volume form equation
\begin{align}
	\lim_{t\to\infty}\frac{1}{t}\int_{0}^{t}\mathrm{d}s\,\operatorname{Tr}\mathsf{A}(s)=\sum_{i=1}^{d}\lambda_{i}
\end{align}

\vspace{.2cm}

\bibliographystyle{iopart-num}
\bibliography{Lyapunov}{} 
\end{document}